%% file: arXiv_PTFlash.tex
\documentclass[preprint, 12pt]{elsarticle}
\makeatletter
\def\ps@pprintTitle{
	\let\@oddhead\@empty
	\let\@evenhead\@empty
	\def\@oddfoot{\centerline{\thepage}}
	\let\@evenfoot\@oddfoot}
\makeatother

\biboptions{sort&compress}
\usepackage{subfigure}
\usepackage{booktabs}
\usepackage{multirow}
\usepackage{hyperref}
\usepackage{threeparttable}
\usepackage{amssymb}
\usepackage{amsmath}
\usepackage{isomath}
\usepackage{stackrel}
\renewcommand{\vec}{\vectorsym}
\newcommand{\mat}{\matrixsym}

\usepackage[linesnumbered, ruled, noend]{algorithm2e}
\usepackage{xcolor}

\SetCommentSty{mycommfont}

\journal{Fuel}

\begin{document}
\begin{frontmatter}

\title{PTFlash : A deep learning framework for isothermal two-phase equilibrium calculations}


\author[sorbonne,ifp]{Jingang Qu}

\author[ifp]{Thibault Faney}

\author[ifp]{Jean-Charles de Hemptinne}

\author[ifp]{Soleiman Yousef}

\author[sorbonne,criteo]{Patrick Gallinari}

\address[sorbonne]{Sorbonne Université, CNRS, ISIR, F-75005 Paris, France}
\address[ifp]{IFPEN, France}
\address[criteo]{Criteo AI Lab, Paris, France}

\begin{abstract}
Phase equilibrium calculations are an essential part of numerical simulations of multi-component multi-phase flow in porous media, accounting for the largest share of the  computational time. In this work, we introduce a GPU-enabled, fast, and parallel framework, \emph{PTFlash}, that vectorizes algorithms required for isothermal two-phase flash calculations using PyTorch, and can facilitate a wide range of downstream applications. In addition, to further accelerate \emph{PTFlash}, we design two task-specific neural networks, one for predicting the stability of given mixtures and the other for providing estimates of the distribution coefficients, which are trained offline and help shorten computation time by sidestepping stability analysis and reducing the number of iterations to reach convergence.

The evaluation of \emph{PTFlash} was conducted on three case studies involving hydrocarbons, $CO_2$ and $N_2$, for which the phase equilibrium was tested over a large range of temperature, pressure and composition conditions, using the Soave-Redlich-Kwong (SRK) equation of state. We compare \emph{PTFlash} with an in-house thermodynamic library, \emph{Carnot}, written in C++ and performing flash calculations one by one on CPU. Results show speed-ups on large scale calculations up to two order of magnitudes, while maintaining perfect precision with the reference solution provided by \emph{Carnot}.
\end{abstract}




\end{frontmatter}

\clearpage
\input{Sections/introduction.tex}
\input{Sections/methods.tex}
\input{Sections/results.tex}
\input{Sections/conclusion.tex}
\clearpage

\appendix
\input{Appendices/srk_eos.tex}
\input{Appendices/trust_region_methods.tex}
\input{Appendices/solve_rr_equation.tex}
\input{Appendices/typical_compositions.tex}
\clearpage
\input{Appendices/vectorized_algorithms.tex}
\clearpage

\bibliographystyle{elsarticle-num} 
\bibliography{reference}

\end{document}

%% file: Sections/introduction.tex
\section{Introduction}   \label{sec:introduction}

Numerical simulation of multi-component multi-phase flow in porous media is an essential tool for many subsurface applications, from reservoir simulation to long term CO2 storage. A core element of the simulator for such applications is to determine the phase distribution of a given fluid mixture at equilibrium, also known as "flash" calculations. Starting with the seminal work of Michelsen \cite{michelsen_isothermal_1982, michelsen_isothermal_1982-1}, researchers have developed robust and efficient algorithms for isothermal two-phase flash calculations. These algorithms have been implemented in the IFPEN thermodynamic C++ library \emph{Carnot}. 

Nonetheless, flash calculations still account for the majority of simulation time in a large range of subsurface applications \cite{wang_non-iterative_1994, belkadi_comparison_2011}. In most simulators, flash calculations are performed for each grid cell at each time step. Moreover, since modern simulators tend to require higher and higher grid resolutions up to billions of grid cells \cite{dogru_next-generation_2009}, the share of computing time due to flash calculations is expected to increase as well. In this context, speeding up flash calculations has drawn increasing research interest.

Some efforts have been made to accelerate flash calculations. \cite{michelsen_simplified_1986, hendriks_reduction_1988, hendriks_application_1992} proposed a reduction method aiming to reduce the number of independent variables by leveraging the sparsity of the binary interaction parameter matrix,resulting in a limited speed-up \cite{belkadi_comparison_2011}. \cite{rasmussen_increasing_2006} introduced the shadow region method using the results of previous time steps to initiate the current one, which assumes that the changes in pressure, temperature, and composition of a given block are small between two adjacent time steps in typical compositional reservoir simulation. \cite{voskov_compositional_2007} presented tie-line based methods, which approximate the results of flash calculations through linear interpolation between existing tie-lines and can be seen as a kind of look-up table. In \cite{gaganis_integrated_2014, gaganis_machine_2012, kashinath_fast_2018, wang_accelerating_2019}, the authors focused on the use of machine learning, which provides a collection of techniques that can effectively discover patterns and regularities in data. They used support vector machine \cite{cortes_support-vector_1995}, relevance vector machine \cite{tipping_sparse_2001} and neural networks \cite{goodfellow_deep_2016} to directly predict equilibrium phases and provide more accurate initial estimates for flash calculations. In \cite{dogru_next-generation_2009, chen_gpu-based_2014}, researchers focused on developing faster parallel linear solvers, with \cite{dogru_next-generation_2009} mentioning specifically that the vectorization of partial equation of state (EOS) related operations would lead to faster execution.

In this work, we introduce \emph{PTFlash}, a framework for isothermal two-phase equilibrium calculations based on the SRK equations of state \cite{soave_equilibrium_1972}. \emph{PTFlash} is built on the deep learning framework PyTorch \cite{paszke_pytorch_2019} and consists in two main elements, namely the vectorization of algorithms and the use of neural networks. First, we perform a complete rewrite of isothermal two-phase flash calculation algorithms of \emph{Carnot} using PyTorch. This enables the systematic vectorization of the complex iterative algorithms implemented in \emph{Carnot}, allowing in turn to efficiently harness modern hardware with the help of, e.g., Advanced Vector Extensions AVX for Intel CPUs \cite{lomont_introduction_2011} and CUDA for Nvidia GPUs \cite{sanders_cuda_2010}. Note that vectorization of complex iterative algorithms with branching is not straightforward and needs specific care. Second, we replace repetitive and time consuming parts of the original algorithms with deep neural networks trained on the exact solution. More specifically, one neural network is used to predict the stability of given mixtures, and another one is used to provide initial estimates for the iterative algorithms. Once well trained, neural networks are seamlessly incorporated into \emph{PTFlash}. These two elements allow \emph{PTFlash} to provide substantial speed-ups compared to \emph{Carnot}, especially so in the context of flow simulations where parallel executions of flash calculations for up to a billion grid cells are needed.

The rest of this article is organized as follows. In Section \ref{sec:flash_calculations}, we introduce the fundamentals of isothermal two-phase flash calculations and present three case studies. In Section \ref{sec:vectorization}, we explain how to efficiently vectorize flash calculations using PyTorch. In Section \ref{sec:deep_learning}, we present two neural networks to speed up calculations. In Section \ref{sec:results}, we demonstrate the attractive speed-up due to vectorization and the introduction of neural networks. Finally, we summarize our work and suggest future research in Section \ref{sec:conclusion}.

%% file: Sections/methods.tex
\section{Isothermal two-phase flash calculations}  \label{sec:flash_calculations}

In this section, we  introduce the essential concepts of isothermal two-phase flash calculations. In the following, without loss of generality, we consider the equilibrium between the liquid and vapor phases.

\subsection{Problem setting}  \label{sec:problem_setting}

We consider a mixture of $N_c$ components. Given pressure ($P$), temperature ($T$) and feed composition ($\vec{z} = (z_1, \ldots, z_{N_c})$), the objective of flash calculations is to determine the system state at equilibrium: single phase or coexistence of two phases. In the latter case, we need to additionally compute the molar fraction of vapor phase $\theta_V$, the composition of the liquid phase $\vec{x}$ and that of the vapor phase $\vec{y}$. These properties are constrained by the following mass balance equations:
\begin{subequations}
	\begin{gather}
		x_i (1 - \theta_V) + y_i \theta_V = z_i, \quad \text{for} \ i = 1, \ldots, N_c              \\
		\sum_{i=1}^{N_c} x_i = \sum_{i=1}^{N_c} y_i = 1
	\end{gather}
	\label{eq:material_balance}
\end{subequations}
In addition, the following equilibrium condition should be satisfied:
\begin{equation}
	\frac{\varphi^L_i(P, T, \vec{x})}{\varphi^V_i(P, T, \vec{y})} = \frac{y_i}{x_i}  \label{eq:eq_cond}
\end{equation}
where the superscripts $L$ and $V$ refer to the liquid and vapor phases, respectively, and $\varphi_i$ is the fugacity coefficient of component $i$, which is a known nonlinear function of $P$, $T$ and the corresponding phase composition. This function depends on an equation of state that relates pressure, temperature and volume. In this work, we use the SRK equation of state \cite{soave_equilibrium_1972} and solve it using an iterative approach rather than the analytical solution of the cubic equation, e.g., the Cardano's formula, which may be subject to numerical errors in certain edge cases \cite{zhi_fallibility_2002}. For more details, see \ref{sec:srk_eos}.

\subsection{Numerical solver}
Equations \ref{eq:material_balance} and \ref{eq:eq_cond} form a non-linear system, which is generally solved in a two-stage procedure. First, we establish the stability of a given mixture via stability analysis (Section \ref{sec:stability_analysis}). If the mixture is stable, only one phase exists at equilibrium. Otherwise, two phases coexist. Second, we determine $\theta_V$, $\vec{x}$ and $\vec{y}$ at equilibrium through phase split calculations (Section \ref{sec:phase_split}).

\subsubsection{Stability analysis} \label{sec:stability_analysis}

A mixture of composition $\vec{z}$ is stable at specified $P$ and $T$ if and only if its total Gibbs energy is at the global minimum, which can be verified through the reduced tangent plane distance \cite{michelsen_isothermal_1982}:
\begin{equation}
	tpd(\vec{w}) = \sum_{i=1}^{N_c} w_i (\ln w_i + \ln \varphi_i(\vec{w}) - \ln z_i - \ln \varphi_i(\vec{z}))  \label{eq:tpd}
\end{equation}
where $\vec{w}$ is a trial phase composition. If $tpd(\vec{w})$ is non-negative for any $\vec{w}$, the mixture is stable. This involves a constrained minimization problem, which is generally reframed as an unconstrained one:
\begin{equation}
	tm(\vec{W}) = \sum_{i=1}^{N_c} W_i (\ln W_i + \ln \varphi_i(\vec{W}) - \ln z_i - \ln \varphi_i(\vec{z}) - 1)  \label{eq:tm}
\end{equation}
where $tm$ is the modified tangent plane distance and $\vec{W}$ is mole numbers. To locate the minima of $tm$, we first use the successive substitution method accelerated by the Dominant Eigenvalue Method (DEM) \cite{orbach_convergence_1971}, which iterates:
\begin{equation}
	\ln W_i^{(k+1)} = \ln z_i + \ln \varphi_i(\vec{z}) - \ln \varphi_i(\vec{W}^{(k)}) \label{eq:ss_analyser}
\end{equation}
It is customary to initiate the minimization with two sets of estimates, that is, vapour-like estimate $W_i = K_i z_i$ and liquid-like estimate $W_i = z_i / K_i$, where $K_i$ is the distribution coefficients, defined as $y_i / x_i$ and initialized via the Wilson approximation \cite{soave_equilibrium_1972}, as follows:
\begin{equation}
	\ln K_i = \ln \left( \frac{P_{c, i}}{P} \right) + 5.373 (1 + \omega_i) \left( 1 - \frac{T_{c, i}}{T} \right) \label{eq:wilson}
\end{equation}
where $T_{c, i}$ and $P_{c, i}$ refer to the critical temperature and pressure of component $i$, respectively, and $\omega_i$ is the acentric factor. 

Once converging to a stationary point (i.e., $\max ( \lvert \partial tm / \partial \vec{W} \rvert ) <$ 1.0$e$-6) or a negative $tm$ is found, successive substitution stops. If this does not happen after a fixed number of iterations (9 in our work), especially in the vicinity of critical points, we resort to a second-order optimization technique, i.e., the trust-region method \cite{hebden_algorithm_1973}, to minimize $tm(\vec{W})$, which we describe in \ref{sec:tr_sa}. In addition, based on the results of stability analysis, we can re-estimate $K_i$ more accurately as $z_i / W^L_i$ if $tm^L < tm^V$ or $W^V_i / z_i$ otherwise, where the superscripts $V$ and $L$ denote the results obtained using the vapor-like and liquid-like estimates, respectively.

\subsubsection{Phase split calculations} \label{sec:phase_split}

Substituting $K_i = y_i / x_i$ into Equations \ref{eq:material_balance} yields the following Rachford Rice equation \cite{rachford_procedure_1952}:
\begin{equation}
	f_{RR}(\theta_V, \vec{K}) = \sum_{i=1}^{N_c} \frac{(K_i - 1) z_i}{1 + (K_i - 1) \theta_V} = 0 \label{eq:RR}
\end{equation}
Given $\vec{K} = (K_1, \ldots, K_{N_c})$, we solve the above equation using the method proposed by \cite{leibovici_new_1992} to get $\theta_V$, which is detailed in \ref{sec:solve_RR}. 

\begin{figure}[h]
	\centering
	\includegraphics[width=0.85\textwidth]{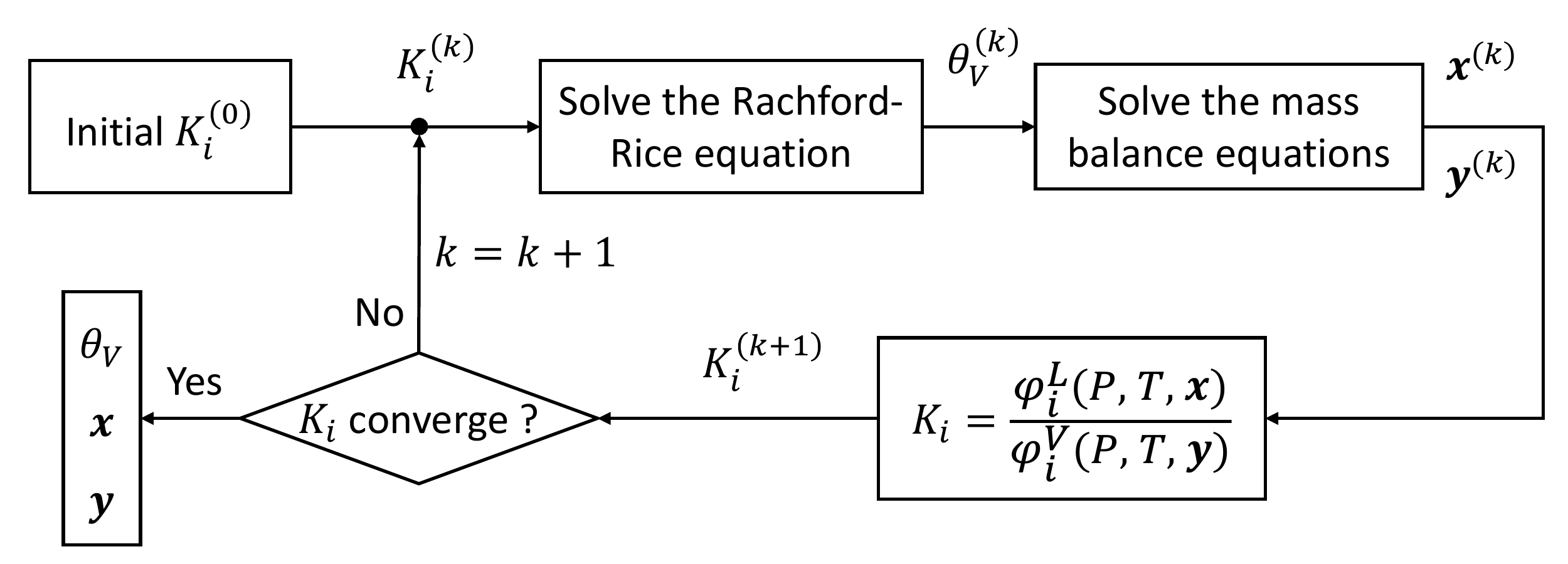}
	\caption[Successive substitution of phase split calculations]{Successive substitution of phase split calculations}
	\label{fig:ss_phase_split}
\end{figure}

To obtain $\theta_V$, $\vec{x}$ and $\vec{y}$ at equilibrium, phase split calculations start with the accelerated successive substitution method, as illustrated in Figure \ref{fig:ss_phase_split}, and the corresponding convergence criterion is $\max( \lvert K_i^{(k+1)} / K_i^{(k)} - 1 \rvert ) < $1.0$e$-8. If successive substitution fails to converge after a few iterations (9 in our work), we use the trust-region method to minimize the reduced Gibbs energy:
\begin{equation}
	G = \sum_{i=1}^{N_c} n_i^L (\ln x_i + \ln \varphi_i^L) + \sum_{i=1}^{N_c} n_i^V (\ln y_i + \ln \varphi_i^V) \label{eq:gibbs}
\end{equation}
where $n_i^L = x_i (1 - \theta_V)$ and $n_i^V = y_i \theta_V$ are the mole numbers of liquid and vapor phases, respectively. The convergence criterion is $\max ( \lvert \partial G / \partial n_i^V \rvert ) <$ 1.0$e$-8. For more details, see \ref{sec:tr_split}.
\subsubsection{Strategy for isothermal two-phase flash calculations}  \label{sec:strategy}

\begin{figure}[h]
	\centering
	\includegraphics[width=0.54\textwidth]{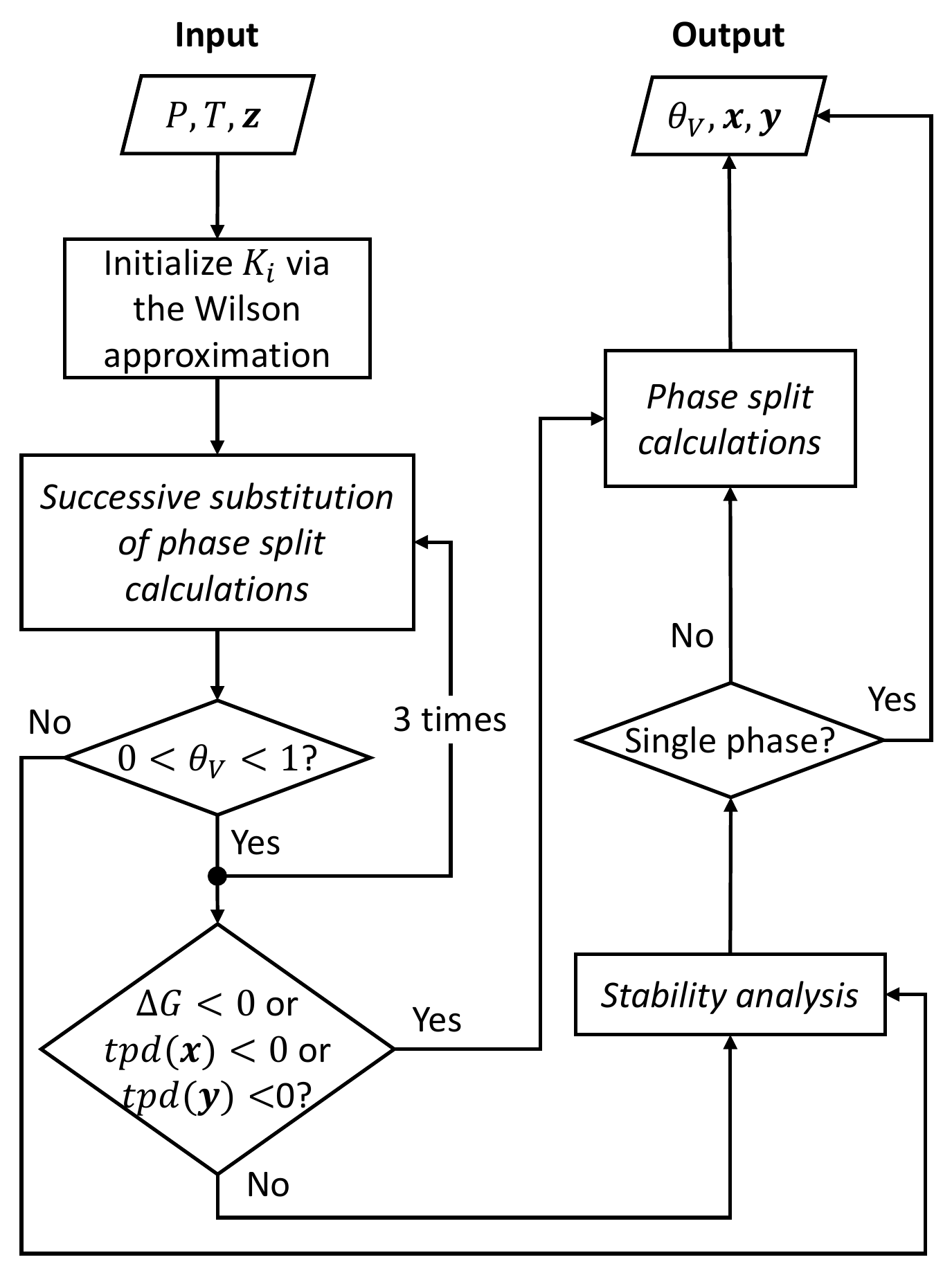}
	\caption[Flowchart of isothermal two-phase flash calculations]{Flowchart of isothermal two-phase flash calculations}
	\label{fig:isothermal_flash}
\end{figure}

We basically adopt the rules of thumb proposed by Michelsen in the book \cite{michelsen_thermodynamic_2004} to implement flash calculations, as shown in Figure \ref{fig:isothermal_flash}. In the flowchart, we first initialize the distribution coefficients $K_i$ using the Wilson approximation. Subsequently, in order to avoid computationally expensive stability analysis, we carry out the successive substitution of phase split calculations 3 times, which will end up with 3 possible cases: (1) $\theta_V$ is out of bounds (0, 1) during iterations. (2) None of $\upDelta G$, $tpd(\vec{x})$ and $tpd(\vec{y})$ are negative, where $tpd(\vec{x})$ and $tpd(\vec{y})$ are reduced tangent plane distances using current vapor and liquid phases as trial phases, and $\upDelta G = \theta_V \times tpd(\vec{x}) + (1 - \theta_V) \times tpd(\vec{y})$. (3) Any of $\upDelta G$, $tpd(\vec{x})$ and $tpd(\vec{y})$ is negative.

For the first two cases, we cannot be sure of the stability of the given mixture, thus continuing with stability analysis. For the third case, we can conclude that the given mixture is unstable, thereby sidestepping stability analysis. Finally, if two phases coexist, we perform phase split calculations to get $\theta_V$, $\vec{x}$ and $\vec{y}$ at equilibrium.
\subsection{Case studies}  \label{sec:case_study}

Here, we introduce three case studies involving hydrocarbons, $CO_2$ and $N_2$, whose properties are shown in Table \ref{tab:properties}. In this work, we only consider the binary interaction parameter (BIP) between $CH_4$ and $CO_2$, which is 0.0882. The BIPs between the others are 0.
\begin{table}[h]
	\centering
	\small
	\caption{Component properties}
	\begin{tabular}{cccc}
		\toprule
		& $P_c$ (MPa) & $T_c$ (K) & $w$ \\
		\midrule
		$CH_4$          & 4.6    & 190.55 & 0.0111 \\
		$C_2H_6$        & 4.875  & 305.43 & 0.097  \\
		$C_3H_8$        & 4.268  & 369.82 & 0.1536 \\
		$n$-$C_4H_{10}$ & 3.796  & 425.16 & 0.2008 \\
		$n$-$C_5H_{12}$ & 3.3332 & 467.15 & 0.2635 \\
		$C_6H_{14}$     & 2.9688 & 507.4  & 0.296  \\
		$C_7H_{16}^+$   & 2.622  & 604.5  & 0.3565 \\
		$CO_2$          & 7.382  & 304.19 & 0.225  \\
		$N_2$           & 3.3944 & 126.25 & 0.039  \\
		\bottomrule
	\end{tabular}
	\label{tab:properties}
\end{table}

\begin{figure}[h]
	\centering
	\subfigure[Phase diagrams]{
		\includegraphics[width=0.45\textwidth]{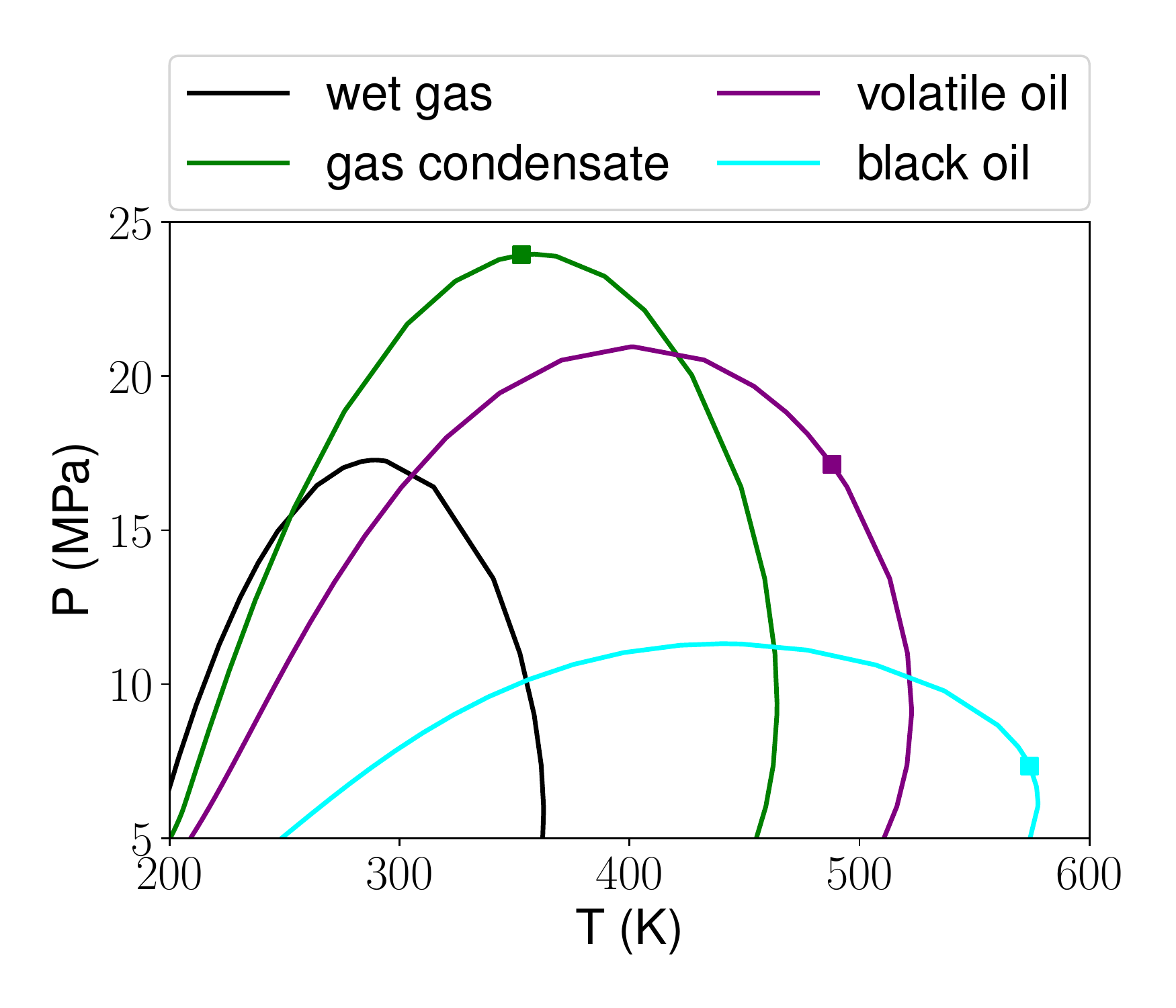}
		\label{fig:envelopes}
	}
	\subfigure[Marginal distribution of $z_i$ for black oil]{
		\includegraphics[width=0.45\textwidth]{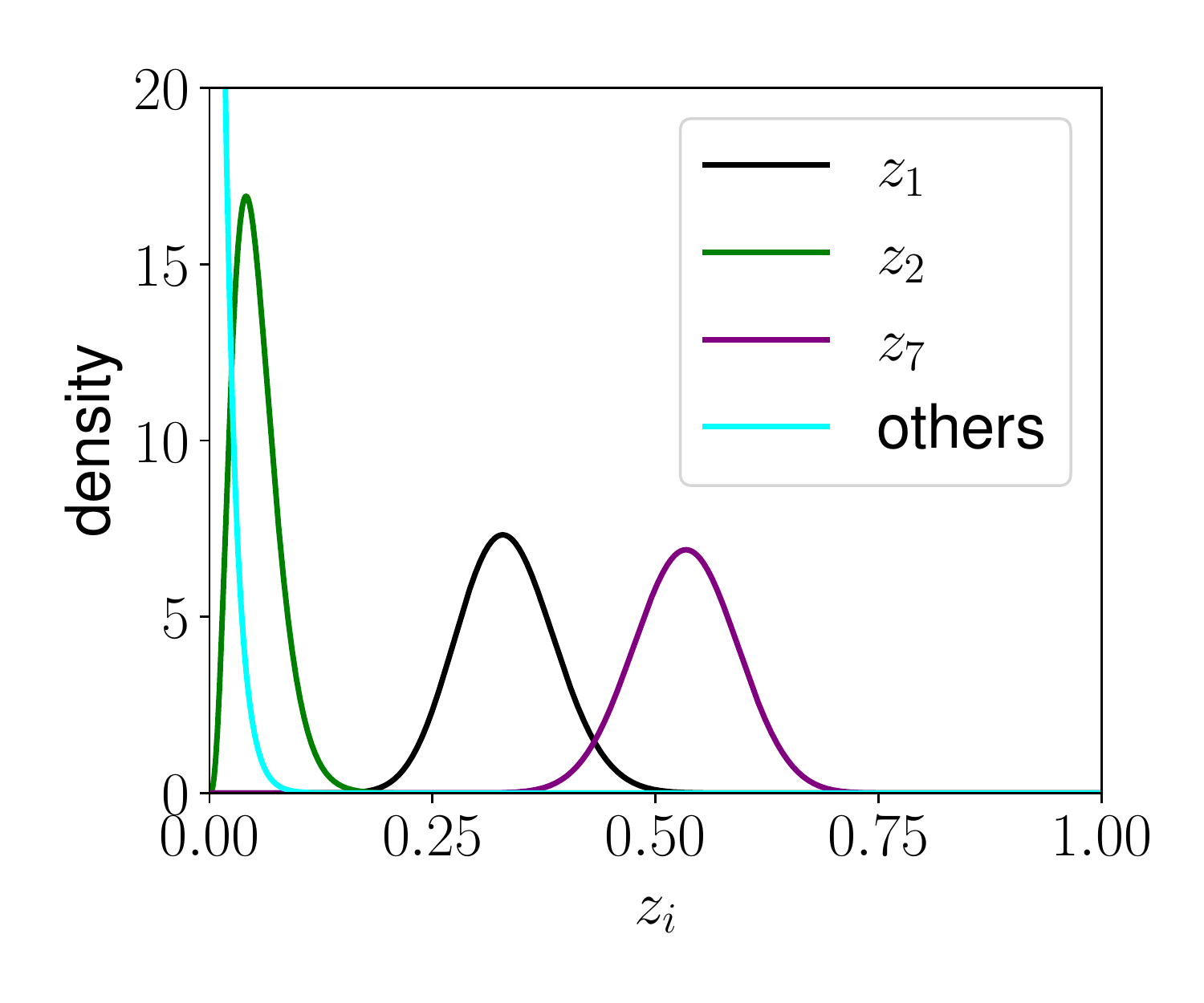}
		\label{fig:black_oil}
	}
	\caption{In Figure (a), the squares on the phase envelopes represent critical points. In Figure (b), $z_1$, $z_2$ and $z_7$ are the molar fractions of $CH_4$, $C_2H_6$ and $C_7H_{16}^+$, respectively.}
\end{figure}

The first case study focuses on a system of two components ($CH_4$ and $C_6H_{14}$), and the second one involves four components ($CH_4$, $C_2H_6$, $C_3H_8$ and $C_4H_{10}$). For these two case studies, the ranges of pressure and temperature are 0.1MPa - 10MPa and 200K - 500K, respectively, and we consider the entire compositional space, i.e., $0 < z_i < 1$ for $i = 1, \ldots, N_c$. The third case study includes all 9 components in Table \ref{tab:properties}. The bounds of pressure and temperature are 5MPa - 25MPa and 200K - 600K, respectively. In addition, from a practical perspective, given that some mixtures do not exist in nature, rather than considering the entire compositional space, we specify four different compositional ranges, as shown in Table \ref{tab:composition_ranges}, each of which represents one of the common reservoir fluid types, namely wet gas, gas condensate, volatile oil, and black oil.  Figure \ref{fig:envelopes} shows phase diagrams of four typical reservoir fluids at fixed compositions (\ref{sec:fixed_compositions}), and we can see that the more heavy hydrocarbons there are, the lower the pressure range of the phase envelope and the less volatile the fluid is.
\begin{table}[htbp]
	\centering
	\small
	\caption{Four fluid types characterized by different compositional ranges}
	\begin{tabular}{ccccc}
		\toprule
		                & Wet gas    & Gas condensate & Volatile oil & Black oil \\
		\midrule
		$CH_4$          & 80\% - 100\% & 60\% - 80\%  & 50\% - 70\%  & 20\% - 40\% \\
		$C_2H_6$        & 2\% - 7\%    & 5\% - 10 \%  & 6\% - 10\%   & 3\% - 6  \% \\
		$C_3H_8$        & $\le 3  \%$  & $\le 4  \%$  & $\le 4.5\%$  & $\le 1.5\%$ \\
		$n$-$C_4H_{10}$ & $\le 2  \%$  & $\le 3  \%$  & $\le 3  \%$  & $\le 1.5\%$ \\
		$n$-$C_5H_{12}$ & $\le 2  \%$  & $\le 2  \%$  & $\le 2  \%$  & $\le 1  \%$ \\
		$C_6H_{14}$     & $\le 2  \%$  & $\le 2  \%$  & $\le 2  \%$  & $\le 2  \%$ \\
		$C_7H_{16}^+$   & $\le 1  \%$  & 5\% - 10 \%  & 10\% - 30\%  & 45\% - 65\% \\
		$CO_2$          & $\le 2  \%$  & $\le 3.5\%$  & $\le 2  \%$  & $\le 0.1\%$ \\
		$N_2$           & $\le 0.5\%$  & $\le 0.5\%$  & $\le 0.5\%$  & $\le 0.5\%$ \\
		\bottomrule
	\end{tabular}
	\label{tab:composition_ranges}
\end{table}

\subsection{Data generation}  \label{sec:generation}

To efficiently sample input data including $P$, $T$ and $\vec{z}$, we first use Latin Hypercube Sampling (LHS) technique to take space-filling samples \cite{mckay_comparison_2000}. Subsequently, for $P$ and $T$, we linearly transform the uniform distribution $\mathcal{U}(0, 1)$ to the expected ranges. For $\vec{z}$ subject to $\sum z_i = 1$, we transform a set of $\mathcal{U}(0, 1)$ into the Dirichlet distribution $Dir(\vec{\alpha})$ whose support is a simplex, as follows:
\begin{subequations}
	\begin{gather}
		x_i \stackrel{i.i.d.}{\sim} \mathcal{U}(0, 1) \ \text{using LHS}    \\
		y_i = \Gamma(\alpha_i, 1).ppf(x_i)                                  \\
		z_i = \frac{y_i}{\sum_{i=1}^{N_c} y_i}
	\end{gather}
	\label{eq:sampling}
\end{subequations}
where $\vec{\alpha} = (\alpha_1, \ldots, \alpha_{N_c})$ is the concentration parameters of the Dirichlet distribution and controls its mode, $\Gamma(\alpha_i, 1)$ is the Gamma distribution, $ppf$ represents the percent-point function, also known as the quantile function, and $\vec{z} = (z_1, \ldots, z_{N_c}) \sim Dir(\vec{\alpha})$. 

For the first two case studies, the concentration parameters are $\vec{\alpha} = \mathbf{1}$, i.e., all-ones vector. For the third case study, we adjust $\vec{\alpha}$ for different fluid types to make the probability of each compositional range as large as possible, as shown in Table \ref{tab:different_alpha}. Figure \ref{fig:black_oil} presents the marginal distribution of $z_i$ for black oil. In summary, we sample $\vec{z}$ using Equations \ref{eq:sampling} with different $\vec{\alpha}$ specified in Table \ref{tab:different_alpha}, and then we single out the acceptable samples located in the compositional ranges defined in Table \ref{tab:composition_ranges}. In the following, unless otherwise specified, four fluid types are always equally represented.
\begin{table}[htbp]
	\centering
	\small
	\caption{Concentration parameters $\vec{\alpha}$ for different fluid types in Table \ref{tab:composition_ranges}}
	\begin{tabular}{ccccc}
		\toprule
		& $\alpha_1$ for $CH_4$ & $\alpha_2$  for $C_2H_6$ & $\alpha_7$ for $C_7H_{16}^+$ & $\alpha_i$ for others \\
		\midrule
		Wet gas        & 100   & 5     & 1     & 1 \\
		Gas condensate & 40    & 5     & 5     & 1 \\
		Volatile oil   & 55    & 8     & 20    & 1 \\
		Black oil      & 25    & 4     & 40    & 1 \\
		\bottomrule
	\end{tabular}
	\label{tab:different_alpha}
\end{table}

Eventually, the samples of $P$, $T$ and $\vec{z}$ are concatenated together to form the complete input data.

\section{Vectorization of isothermal two-phase flash calculations}  \label{sec:vectorization}
We vectorize the isothermal two-phase flash so that it takes as inputs $\vec{P}=(P_1, \cdots, P_n)$, $\vec{T} = (T_1, \cdots, T_n)$ and $\textbf{z} = (\vec{z}_1, \cdots, \vec{z}_n)$, where $\vec{P}$ and $\vec{T}$ are vectors, $\textbf{z}$ is a matrix, and $n$ denotes the number of samples processed concurrently and is often referred to as the batch dimension. 

In recent years, Automatic Vectorization (AV) has emerged and developed \footnote{At the time of writing, PyTorch team released a fledgling library, \emph{functorch}, which takes inspiration from JAX and supports Automatic Vectorization.}, e.g., JAX \cite{bradbury_jax_2018}, which can automatically vectorize a function through the batching transformation that adds a batch dimension to its input. In this way, the vectorized function can process a batch of inputs simultaneously rather than processing them one by one in a loop. However, AV comes at the expense of performance to some extent and is slower than well-designed manual vectorization, which vectorizes a function by carefully revamping its internal operations to accommodate to a batch of inputs. For example, matrix-vector products for a batch of vectors can be directly replaced with a matrix-matrix product. In addition, flash calculations have an iterative nature and complicated control flow, which is likely to result in the failure of AV. Consequently, for finer-grained control, more flexibility, and better performance, we manually vectorize all algorithms involved in flash calculations, including the solution of the SRK equation of state and the Rachford-Rice equation, stability analysis and phase split calculations.

To achieve efficient vectorization, one difficulty is asynchronous convergence, that is, for each algorithm, the number of iterations required to reach convergence generally varies for different samples, which hinders vectorization and parallelism. To alleviate this problem, we design a general-purpose paradigm, \textit{synchronizer}, to save converged results in time at the end of each iteration and then remove the corresponding samples in order not to waste computational resources on them in the following iterations, which is achieved by leveraging a one-dimensional Boolean mask encapsulating convergence information to efficiently access data in vectors and matrices, as follows:
\begin{subequations}
	\begin{gather}
		\mat{X}^{(k+1)} \leftarrow f(\mat{X}^{(k)})                 \\
		\mathrm{Save} \quad \mat{X}^{(k+1)}[\mathrm{mask}] \quad \mathrm{to} \quad \widetilde{\mat{X}}  \\
		\mat{X}^{(k+1)} \leftarrow \mat{X}^{(k+1)}[\sim \mathrm{mask}]  \\
		k \leftarrow k + 1
	\end{gather}
\end{subequations}
where $k$ is the number of iterations, $f(\mat{X})$ is a vectorized iterated function taking as input $\mat{X} \in \mathbb{R}^{n \times m}$ ($n$ is the batch dimension, i.e., number of samples, and $m$ is the dimension of $X$), $\widetilde{\mat{X}}$ is a placeholder matrix used to save converged results, mask is a Boolean vector where True means convergence, and $\sim$ denotes the logical NOT operator. The number of unconverged samples gradually decreases as a result of incremental convergence. For the full version of \textit{synchronizer}, refer to \ref{sec:synchronizer}. We can use \textit{synchronizer} to wrap and vectorize any iterative algorithm. For instance, we illustrate how to perform vectorized stability analysis in \ref{sec:vectorized_sa}.

The efficiency of \textit{synchronizer} may be questioned because previously converged samples are still waiting for unconverged ones before moving to the next step. This is true, but the situation is not as pessimistic since we try to shorten the waiting time as much as possible. For example, if successive substitution fails to converge quickly, we immediately use the trust-region method. In any case, the delay caused by waiting is insignificant compared to the acceleration due to vectorization. Furthermore, we leverage neural networks to provide more accurate initial estimate $\mat{X}^{(0)}$ so that all samples converge as simultaneously as possible, thereby reducing asynchrony, which we will present in Section \ref{sec:deep_learning}.

Once all algorithms are well vectorized, another problem is how to globally coordinate different subroutines. To this end, we add barrier synchronization to the entry points of stability analysis and phase split calculations in Figure \ref{fig:isothermal_flash}, which can avoid any subroutine connected to it proceeding further until all others terminate and arrive at this barrier.

We also optimized the code using TorchScript \cite{paszke_pytorch_2019}, allowing for more efficient execution through algebraic peephole optimizations and fusion of some operations, and more practical asynchronous parallelism without the Python global interpreter lock \cite{van_rossum_python_2011}, whereby vapor-like and liquid-like estimates are dealt with in parallel in stability analysis.

\section{Acceleration of flash calculations using neural networks}  \label{sec:deep_learning}

To further accelerate flash calculations, we create and train two task-specific neural networks, classifier and initializer. The classifier is used to predict the probability $p$ that a given mixture is stable, i.e., $p$ = classifier$(P, T, \vec{z})$, which involves a binary classification problem. It can predict the stability of most samples, thereby bypassing stability analysis and saving time. The initializer is able to initialize $K_i$ more accurately than the Wilson approximation, i.e., $\ln K_i$ = initializer$(P, T, \vec{z})$, which relates to a regression problem. It can reduce the number of iterations required to reach convergence and alleviate the asynchronous convergence we introduced before. Note that the hyper-parameters of neural networks presented below, e.g., the number of units and layers, are dedicated to the case study containing 9 components. Nonetheless, the basic architecture of neural networks and the training methods can be generalized to any case.

\subsection{Classifier}

\subsubsection{Architecture}
The classifier has 3 hidden layers with 32 neurons and using the SiLU activation function \cite{hendrycks_gaussian_2016, elfwing_sigmoid-weighted_2018, ramachandran_swish_2017}. The output layer has only one neuron and uses the sigmoid activation function compressing a real number to the range (0, 1). The input $\vec{x}$ consists of $P$, $T$ and $\vec{z}$, and the output is the probability $p$ that a given mixture is stable. The scaling layer standardizes the inputs as $(\vec{x} - \vec{u}) / \vec{s}$, where $\vec{u}$ and $\vec{s}$ are the mean and standard deviation of $\vec{x}$ over the training set. To train the classifier, we use the binary cross-entropy (bce), which is the de-facto loss function for binary classification problems and defined as:
\begin{equation}
	\text{bce}(y, p) = y \ln p + (1 - y) \ln (1 - p)
\end{equation}
where $y$ is either 0 for unstable mixtures or 1 for stable ones.

\begin{figure}[htbp]
	\centering
	\subfigure[The architecture of classifier]{
		\includegraphics[width=0.45\textwidth]{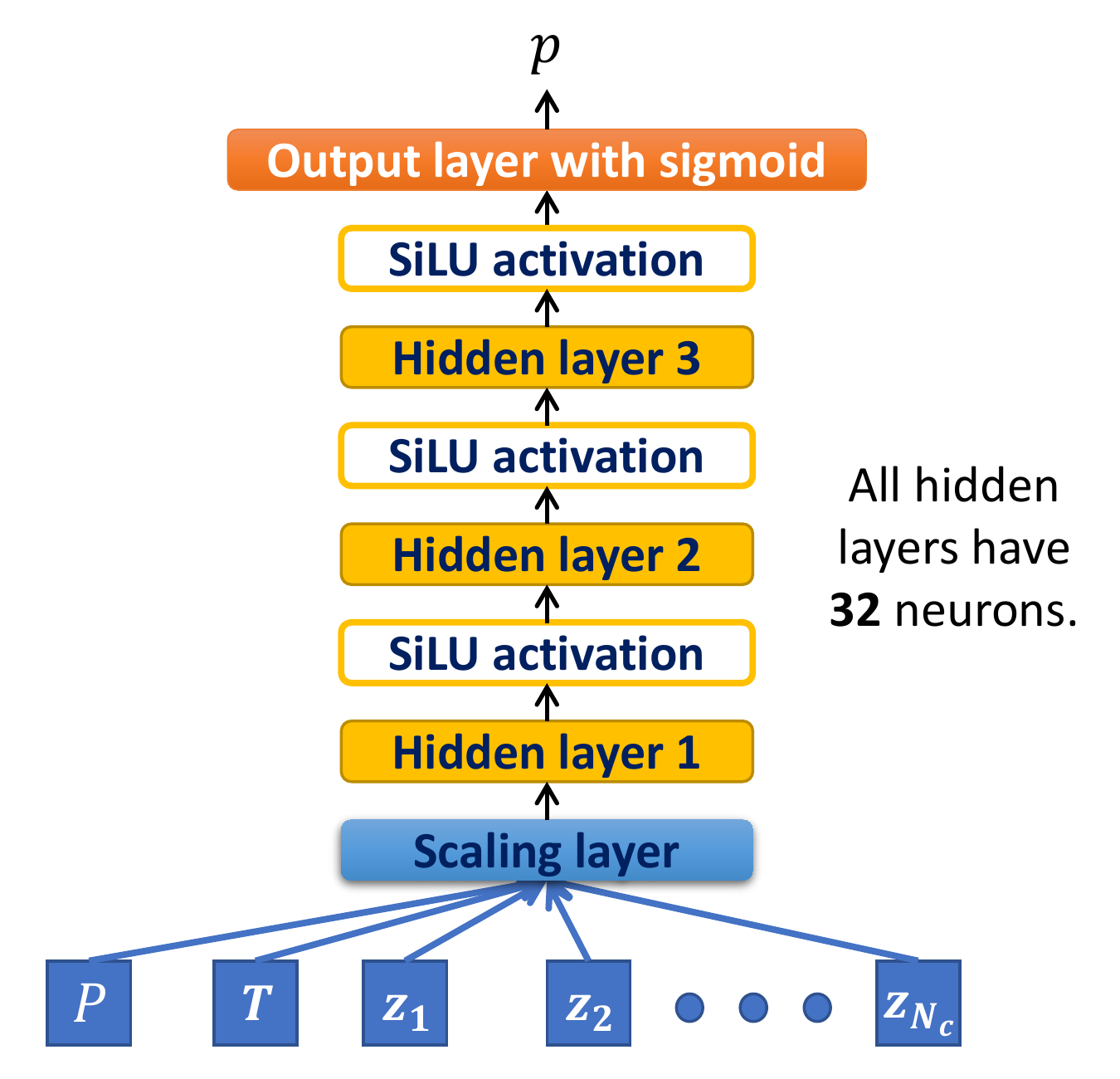}
		\label{fig:classifer}
	}
	\subfigure[Tuning hyper-parameters of classifier]{
		\includegraphics[width=0.48\textwidth]{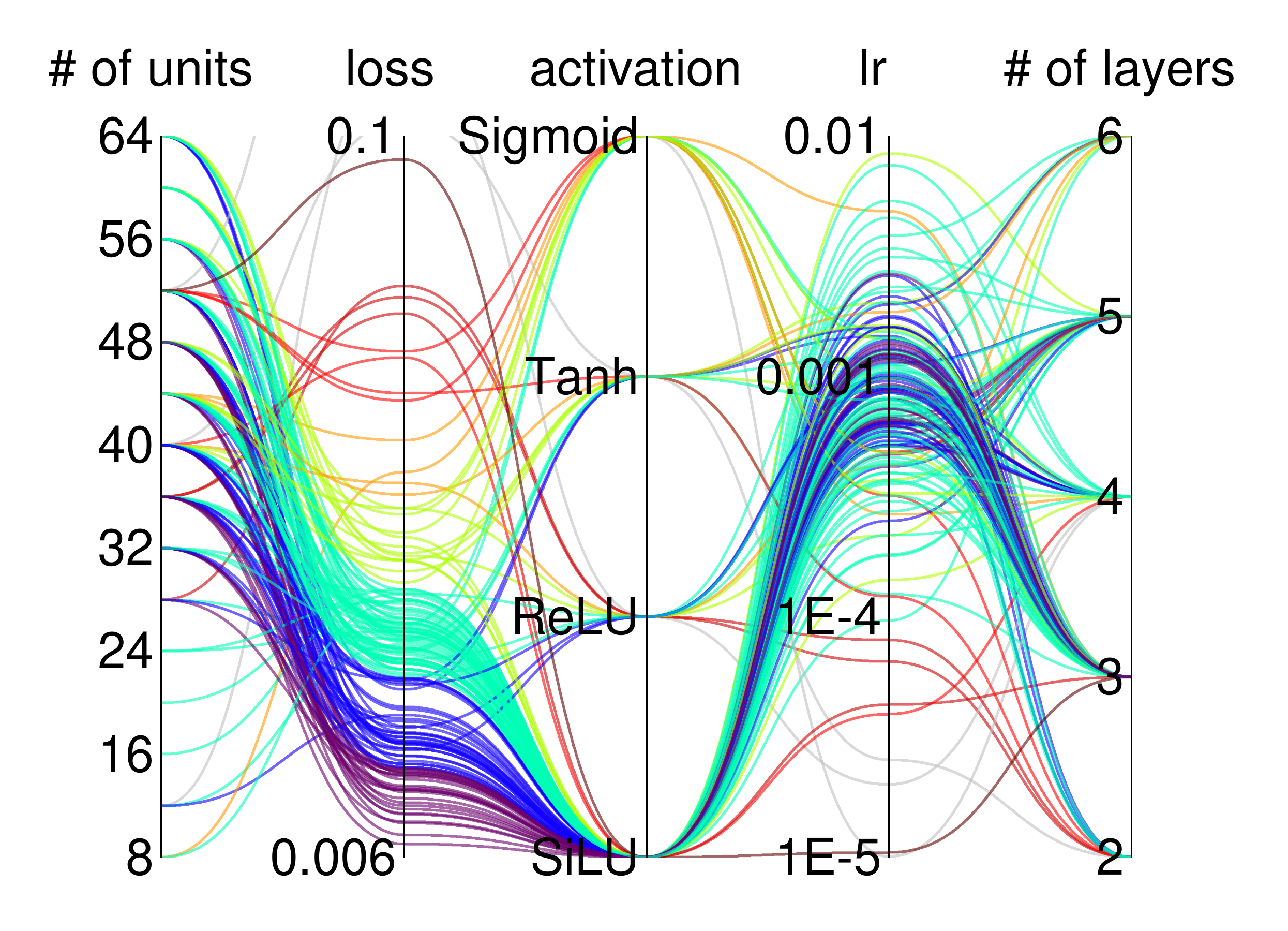}
		\label{fig:hypertuning_classifier}
	}
	\caption{Figure (a) shows the architecture of the classifier. Figure (b) is a parallel coordinates plot used to visualize the results of tuning hyper-parameters of the classifier, where lr stands for learning rate. The colors of lines are mapped to the value of the loss.}
\end{figure}

The architecture of the classifier is obtained by tuning hyper-parameters using Tree-Structured Parzen Estimator optimization algorithm \cite{bergstra_algorithms_2011} with Asynchronous Successive Halving algorithm \cite{li_system_2020} as an auxiliary tool to early stop less promising trials. We create a dataset containing 100,000 samples (80\% for training and 20\% for validation), and then tune the hyper-parameters of the classifier with 150 trials to minimize the loss on the validation set (we use Adam \cite{kingma_adam_2014} as optimizer and the batch size is 512), as shown in Figure \ref{fig:hypertuning_classifier}. We can see that SiLU largely outperforms other activation functions.

\subsubsection{Training}
\begin{figure}[htbp]
	\centering
	\subfigure[Cyclic learning rate of the classifier]{
		\includegraphics[width=0.45\textwidth]{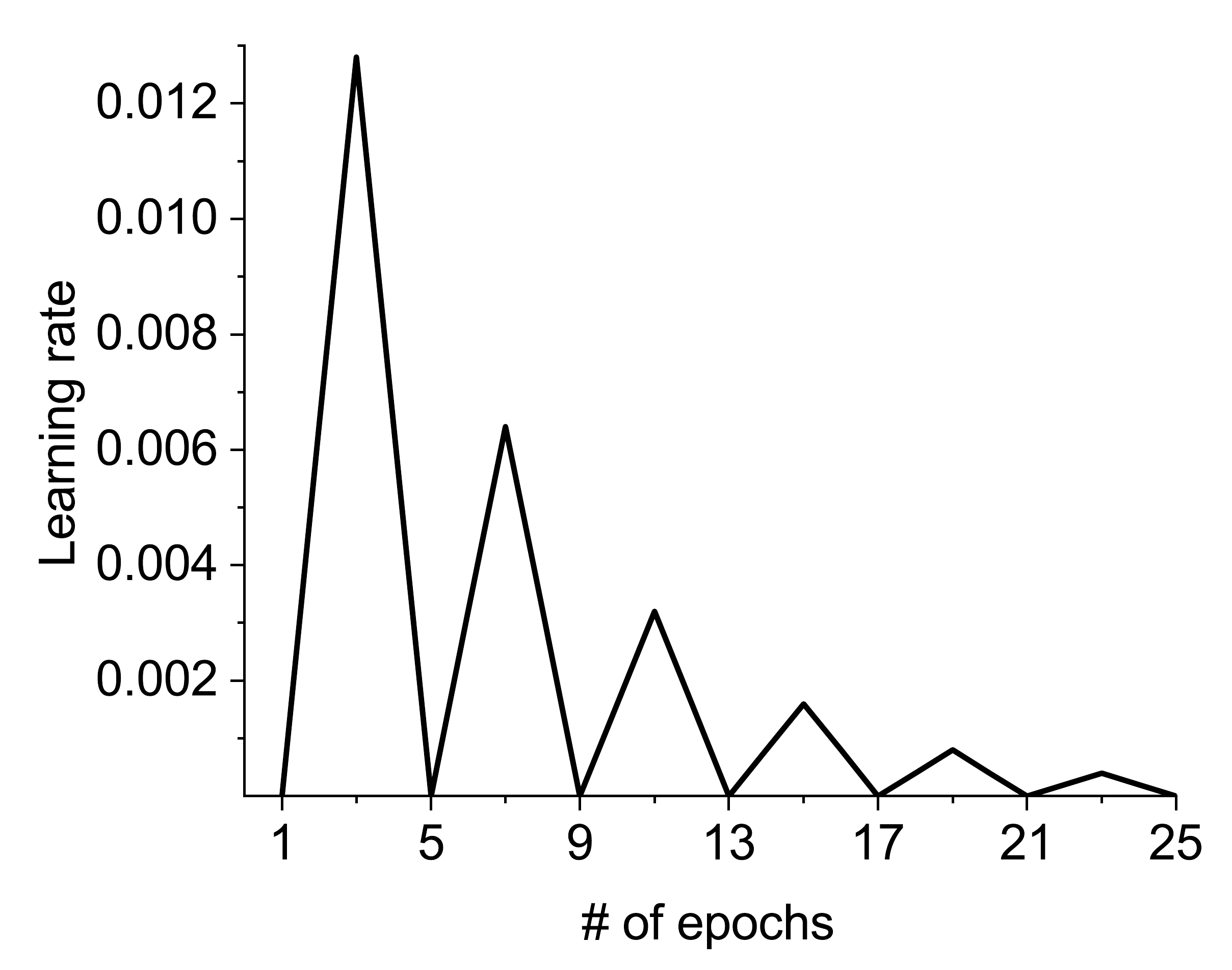}
		\label{fig:lr_schedule}
	}
	\subfigure[Prediction of the classifier for volatile oil]{
		\includegraphics[width=0.45\textwidth]{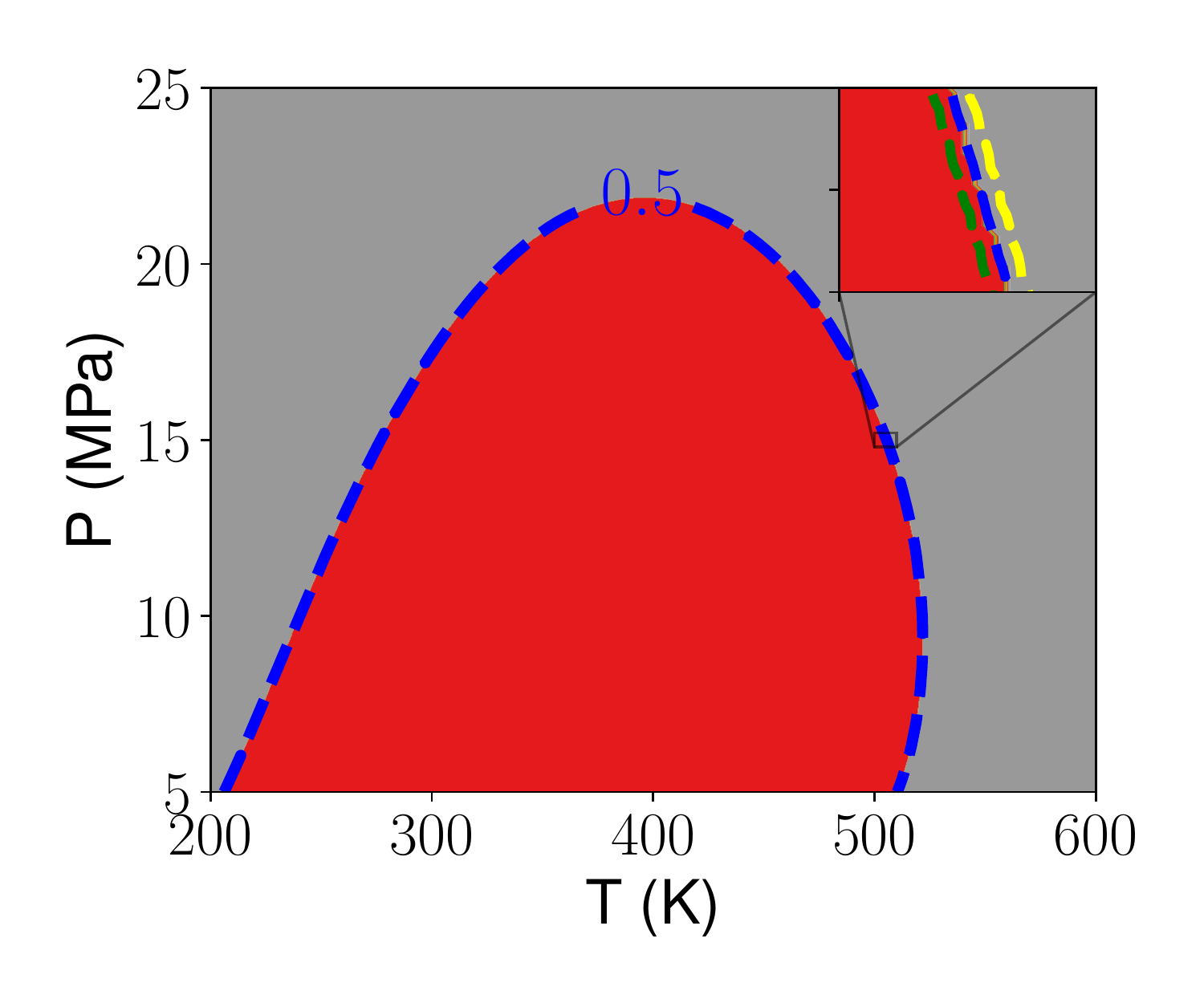}
		\label{fig:classifier_contour}
	}
	\caption{Figure (a) shows how the learning rate varies cyclically. Figure (b) illustrates the contours of probabilities predicted by the classifier for volatile oil at fixed composition. The red and gray correspond to the two-phase and monophasic regions, respectively.}
\end{figure}

We first generate one million samples in the way described in Section \ref{sec:case_study}, and then feed them to \emph{PTFlash} to determine stability (no need for phase split calculations), which takes about 2 seconds. Subsequently, these samples are divided into the training (70\%), validation (15\%) and test (15\%) sets. To train the classifier, we set the batch size to 512 and use Adam with Triangular Cyclic Learning Rate (CLR) \cite{smith_no_2015, smith_cyclical_2017}, which periodically increases and decreases the learning rate during training, as shown in Figure \ref{fig:lr_schedule}. CLR helps neural networks achieve superb performance using fewer epochs and less time \cite{smith_super-convergence_2019}. Early stopping is also used to avoid overfitting \cite{prechelt_early_1998}. The total training time is about 5 minutes using Nvidia RTX 3080. The final performance of the classifier on the test set is bce = 0.002 and accuracy = 99.93\%. For a more intuitive understanding of performance, Figure \ref{fig:classifier_contour} shows the contours of probabilities predicted by the classifier, where the blue contour of $p=0.5$ basically coincides with the phase envelope. In the zoomed inset, the additional green and yellow contours correspond to $p$=0.02 and 0.98, respectively, which are quite close to that of $p=0.5$. This means that the prediction of the classifier is very accurate.

\subsection{Initializer}

\subsubsection{Architecture}
\begin{figure}[htbp]
	\centering
	\includegraphics[width=0.85\textwidth]{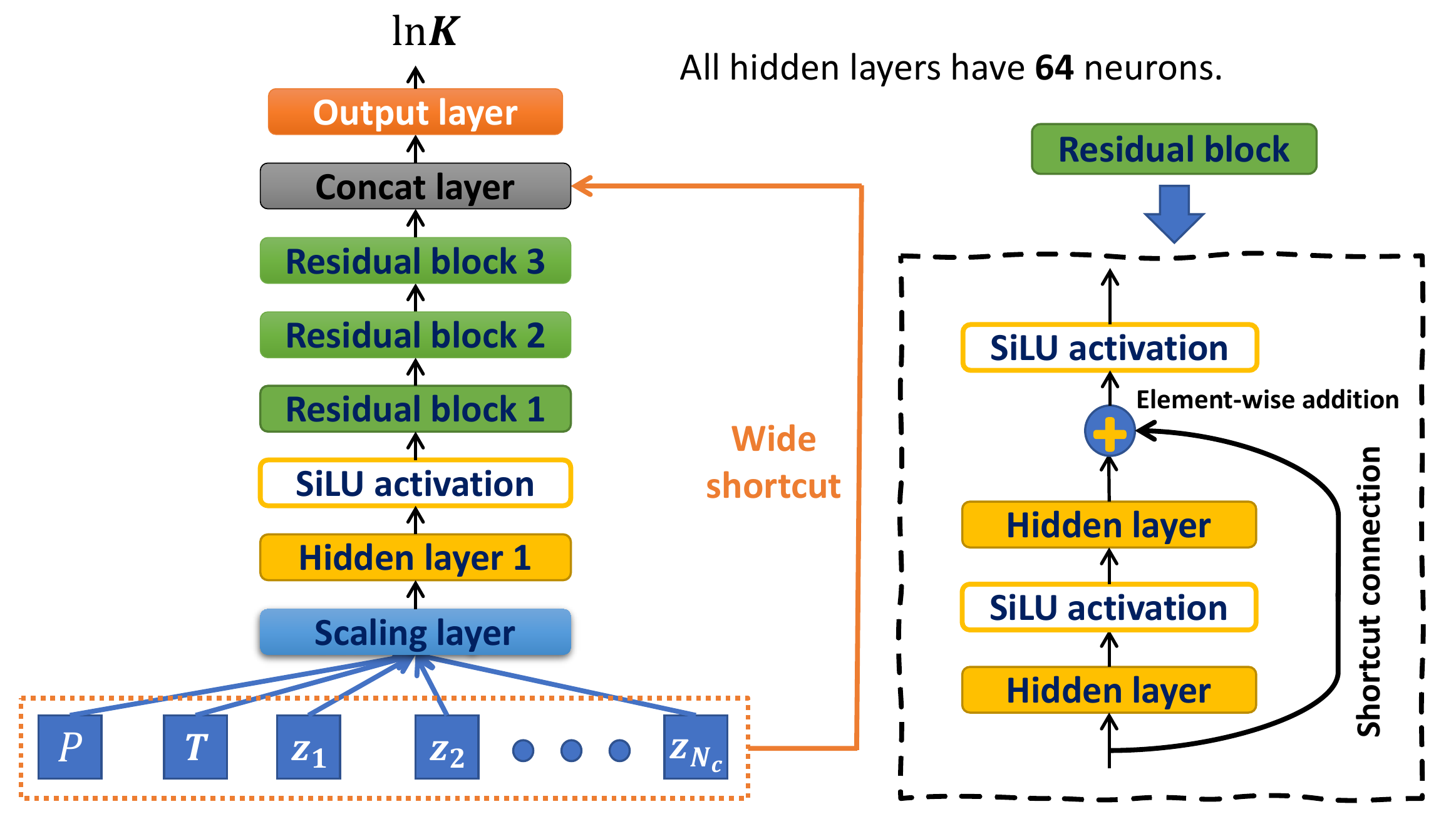}
	\caption{The architecture of initializer}
	\label{fig:initializer}
\end{figure}

The initializer has 1 hidden layer and 3 residual blocks, as shown in Figure \ref{fig:initializer}. Each residual block has 2 hidden layers and a shortcut connection adding the input of the first hidden layer to the output of the second \cite{he_deep_2016}. All hidden layers have 64 neurons and use the SiLU activation function. The output layer has $N_c$ neurons without activation function. The wide shortcut, proposed in \cite{cheng_wide_2016}, enables neural networks to directly learn simple rules via it besides deep patterns through hidden layers, which is motivated by the fact that the inputs, such as $P$ and $T$, are directly involved in the calculation of $K_i$. The concat layer concatenates the input layer and the outputs of the last residual block (the concatenation means putting two matrices $A \in \mathbb{R}^{d_1 \times d_2}$ and $B \in \mathbb{R}^{d_1 \times d_3}$ together to form a new one $C \in \mathbb{R}^{d_1 \times (d_2+d_3)}$). In addition, the targets of the initializer are $\ln K_i$ instead of $K_i$, since $K_i$ varies in different orders of magnitude, which hampers the training of the initializer, whereas $\ln K_i$ does not.

We found that the convergence of phase split calculations is robuster if $K_i$ predicted by the initializer can lead to more accurate values of the vapor fraction $\theta_V$, especially around critical points where calculations are quite sensitive to initial $K_i$ and prone to degenerate into trivial solutions. As a consequence, the loss function used to train the initializer consists of two parts, one is the mean absolute error (mae) in terms of $K_i$ and the other is mae in terms of $\theta_V$, as follows:
\begin{subequations}
	\begin{gather}
		\text{mae}(\ln \vec{K}, \ln \vec{\hat{K}}) = \sum_{i=1}^{N_c} \lvert \ln K_i - \ln \hat{K}_i \rvert  \\
		\text{mae}(\theta_V, \hat{\theta}_V) = \lvert \theta_V - \hat{\theta}_V \rvert
	\end{gather}
\end{subequations}
where $\ln \vec{K}$ is the ground truth, $\ln \vec{\hat{K}}$ is the prediction of the initializer, $\theta_V$ is the vapor fraction at equilibrium, and $\hat{\theta}_V$ is obtained by solving the Rice-Rachford equation given $\vec{z}$ and the prediction $\vec{\hat{K}}$. 

\subsubsection{Training}  \label{sec:training_initializer}
We generate one million samples on the two-phase region ($K_i$ is not available at the monophasic region), which are divided into the training (70\%), validation (15\%) and test (15\%) sets. The training of the initializer is carried out in two stages. First, we train it to minimize $\text{mae}(\ln \vec{K}, \ln \vec{\hat{K}})$, using Adam with CLR and setting the batch size to 512. Second, after the above training, we further train it to minimize $\text{mae}(\ln \vec{K}, \ln \vec{\hat{K}}) + \text{mae}(\theta_V, \hat{\theta}_V)$, using Adam with a small learning rate 1.0$e$-5. Here, $\partial \hat{\theta}_V / \partial \vec{\hat{K}}$ is required during backpropagation and can be simply computed via PyTorch's automatic differentiation, which, however, differentiates through all the unrolled iterations. Instead, we make use of the implicit function theorem \cite{krantz_implicit_2002}, as follows:
\begin{equation}
	\partial \hat{\theta}_V / \partial \vec{\hat{K}} = -[\partial_{\theta_V} f_{RR}(\hat{\theta}_V, \vec{\hat{K}})]^{-1} \partial_{\vec{K}} f_{RR}(\hat{\theta}_V, \vec{\hat{K}})
\end{equation}
In this way, we efficiently obtain $\partial \hat{\theta}_V / \partial \vec{\hat{K}}$ by only using the derivative information at the solution point of the Rachford-Rice equation $(\hat{\theta}_V, \vec{\hat{K}})$, avoiding differentiating through iterations.

Eventually, the performance of the initializer on the test set is mae = 9.66$e$-4 in terms of $\ln K_i$ and mae = 1.86$e$-3 in terms of $K_i$.

\subsection{Strategy for accelerating flash calculations using neural networks}
\begin{figure}[h]
	\centering
	\includegraphics[width=0.55\textwidth]{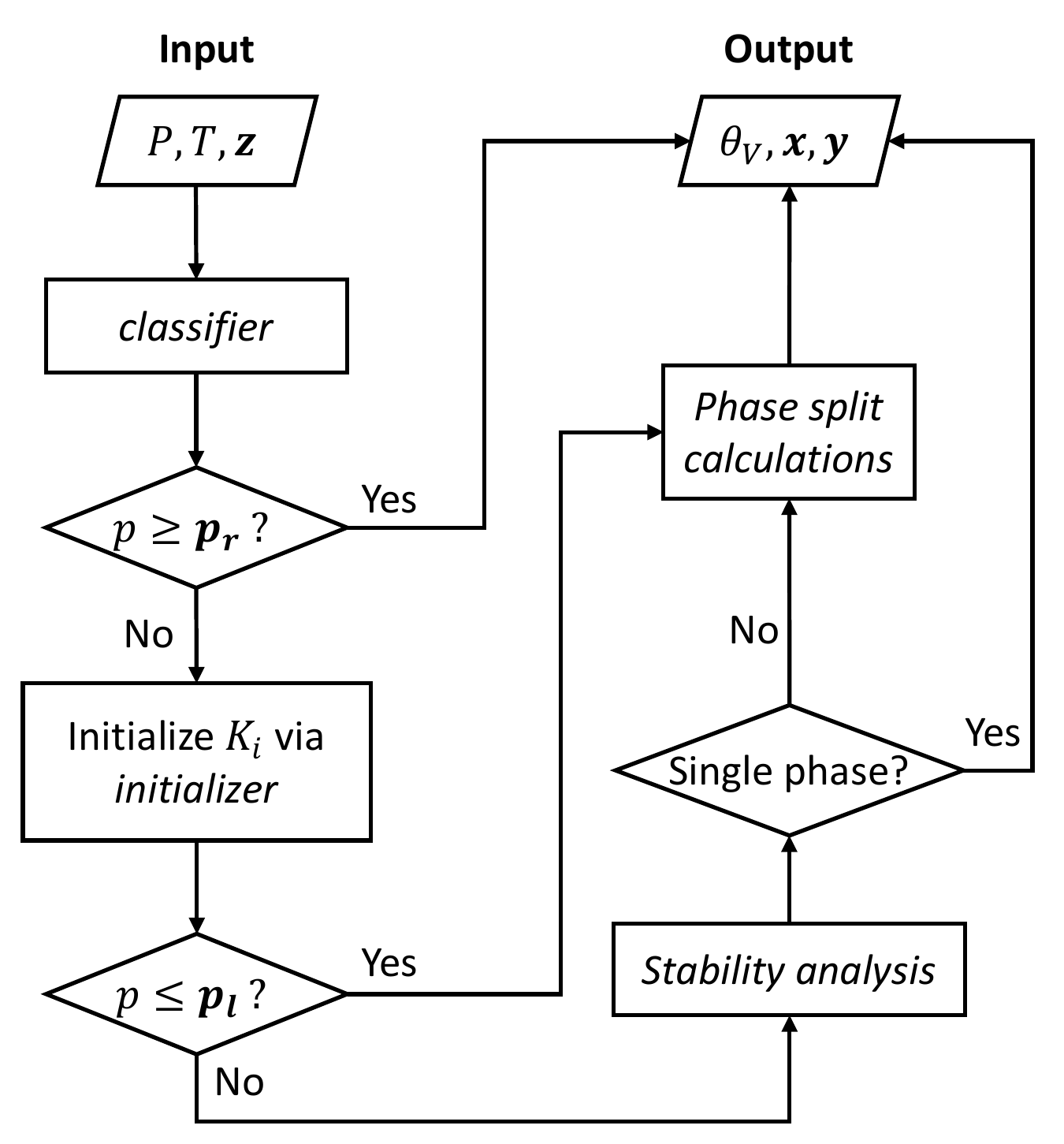}
	\caption[Acceleration of flash calculations using neural networks]{Acceleration of flash calculations using neural networks}
	\label{fig:dl_vectorized_flash}
\end{figure}

As shown in Figure \ref{fig:dl_vectorized_flash}, given $P$, $T$ and $\vec{z}$, we first use the classifier to predict $p$. Next, based on two predefined thresholds, $p_l$ and $p_r$, satisfying $p_l \le p_r$, the given mixture is thought of as unstable if $p \le p_l$ or stable if $p \ge p_r$. If $p_l < p < p_r$, we will use stability analysis to avoid unexpected errors. Here, we can adjust $p_l$ and $p_r$ to trade reliability for speed. In general, less errors occur with smaller $p_l$ and greater $p_r$, but probably taking more time on stability analysis, and vice versa. A special case is $p_l = p_r = p_c$, where $p_c$ could be a well-calibrated probability or simply set to 0.5, which means that we completely trust the classifier (i.e., stable if $p \ge p_c$ or unstable otherwise), and no extra stability analysis is required. For the initializer, it serves both stability analysis when $p_l < p < p_r$ and phase split calculations. 

Neural networks can also be used individually. If only the classifier is available, one may initialize $K_i$ via the Wilson approximation rather than the initializer in Figure \ref{fig:dl_vectorized_flash}. If only the initializer is available, one may use it to initialize $K_i$ in Figure \ref{fig:isothermal_flash}.

%% file: Sections/results.tex
\section{Results} \label{sec:results}

In this section, we will compare our proposed framework for vectorized flash calculations, \textit{PTFlash}, with \textit{Carnot}, an in-house thermodynamic library developed by IFP Energies Nouvelles and based on C++. \textit{Carnot} performs isothermal two-phase flash calculations in the manner shown in Figure \ref{fig:isothermal_flash}, but can only handle samples one at a time on CPU. Regarding the hardware, CPU is Intel 11700F, whose max turbo frequency can reach 4.9GHz, and GPU is NVIDIA RTX 3080 featuring 8704 CUDA cores and 10G memory. Note that since using multiple cores renders the frequency quite unstable due to heat accumulation, we only use one core of CPU so that the frequency can be stabilized at 4.5GHz, which allows for a consistent criterion for measuring the execution time.

\textit{PTFlash} and \textit{Carnot} gave identical results (coincidence to 9 decimal places under double-precision floating-point format) because they use exactly the same convergence criteria for all iterative algorithms. In the following, we will focus on comparing their speeds.

\subsection{Vectorized flash calculations}

\begin{figure}[h]
	\centering
	\subfigure[Mixture of $CH_4$ and $C_6H_{14}$]{
		\includegraphics[width=0.47\textwidth]{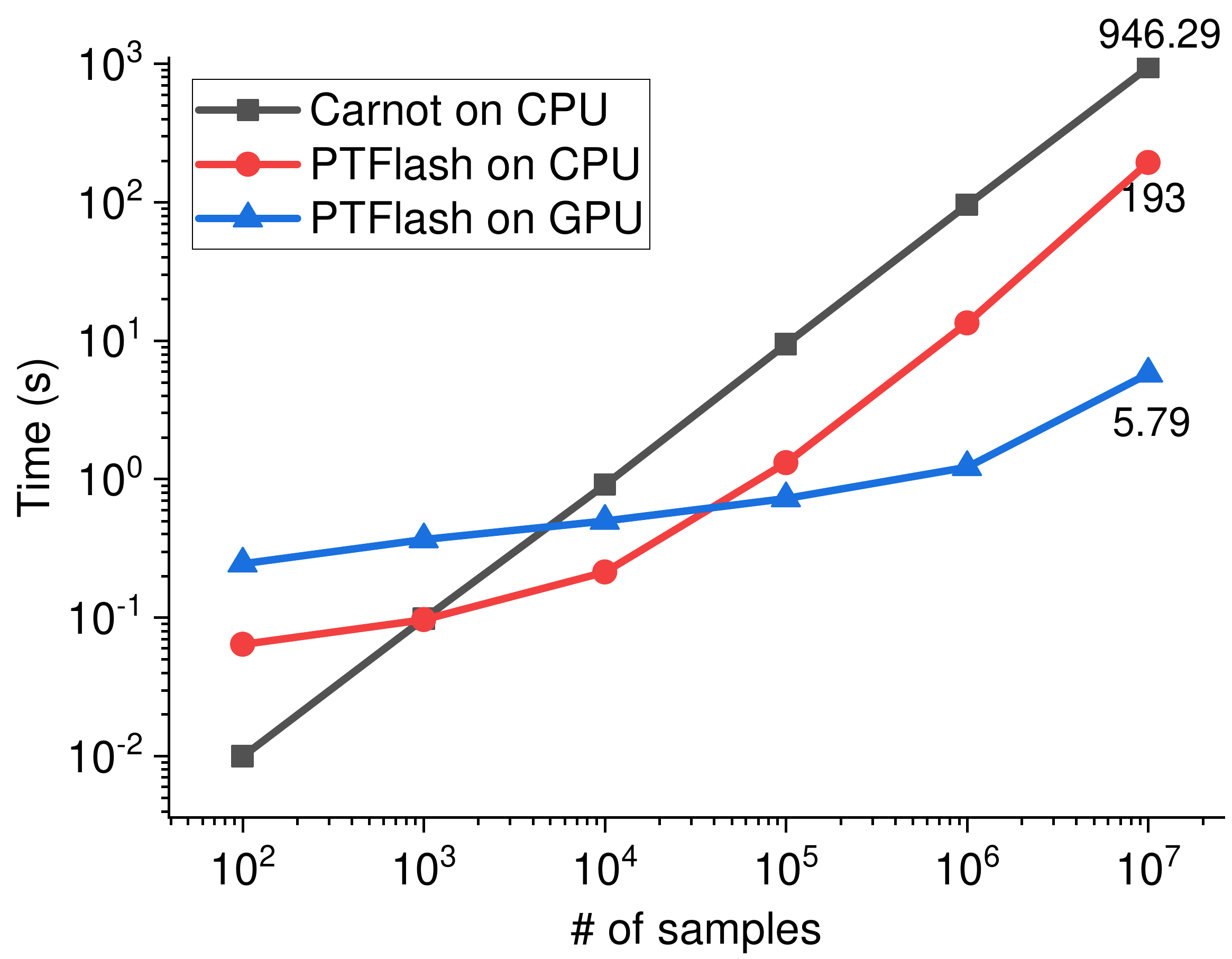}
		\label{fig:compare_2_components}
	}
	\subfigure[Mixture of 4 components]{
		\includegraphics[width=0.47\textwidth]{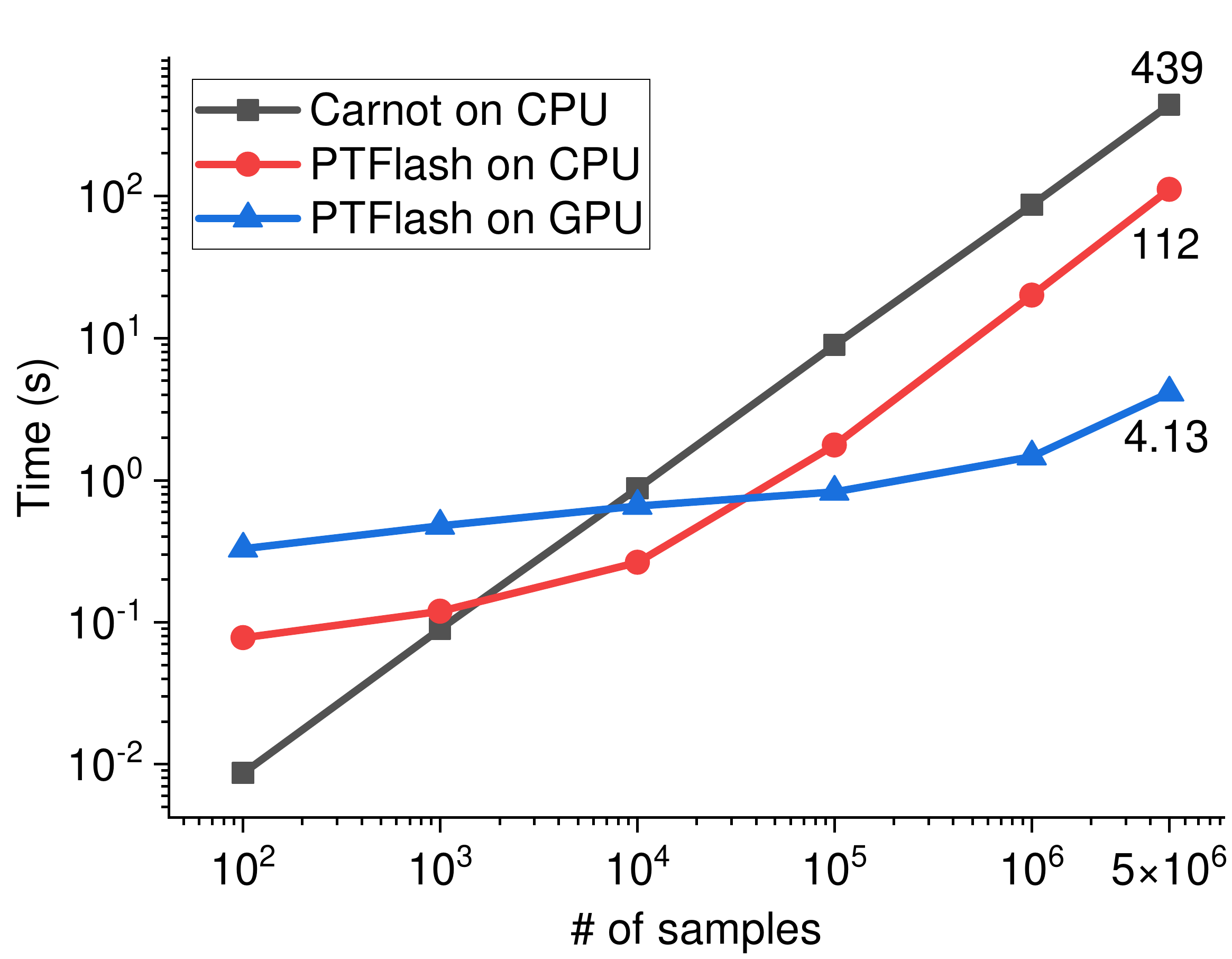}
		\label{fig:compare_4_components}
	}
	\subfigure[Mixture of 9 components]{
		\includegraphics[width=0.5\textwidth]{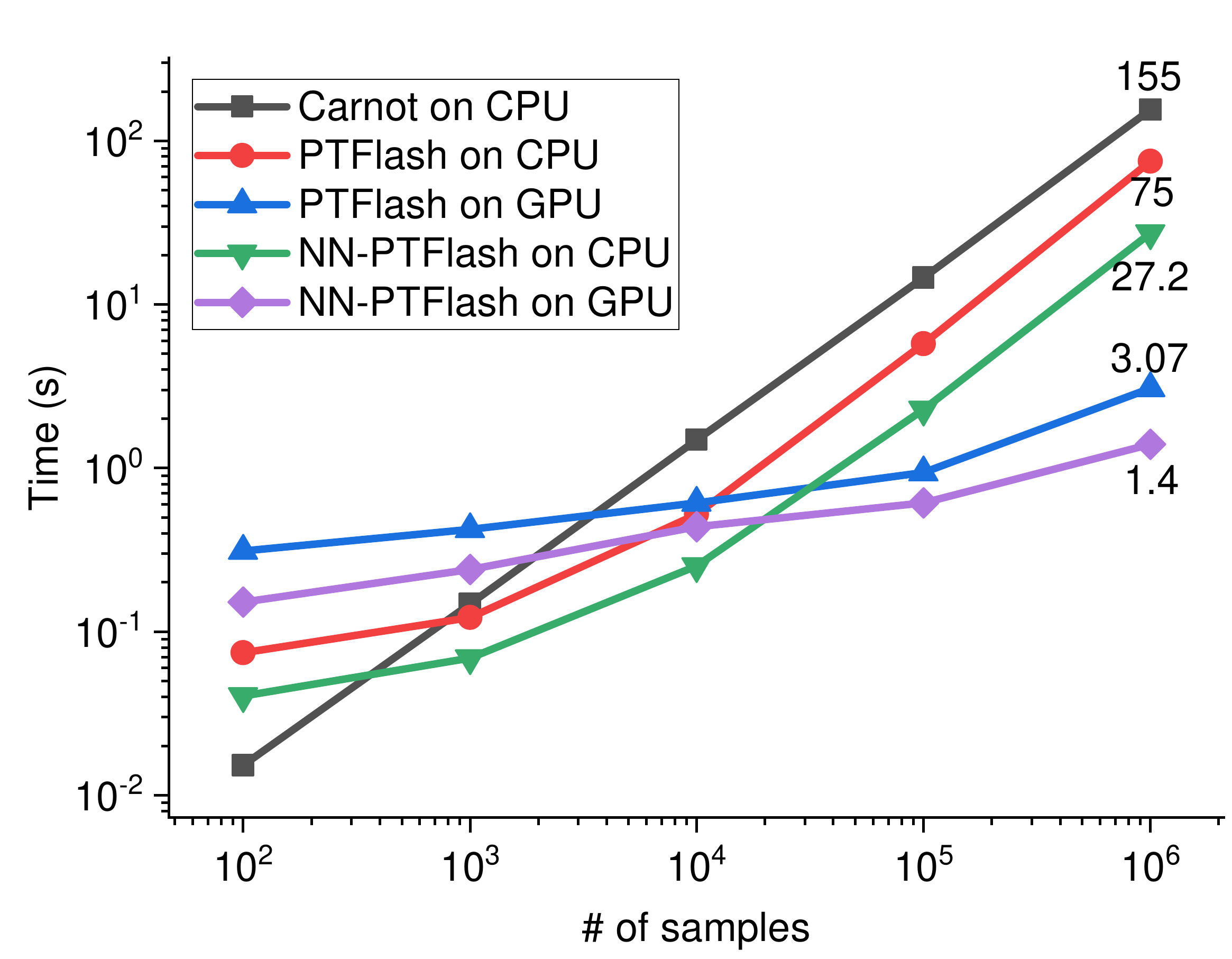}
		\label{fig:compare_9_components}
	}
	\caption{Comparison between \emph{PTFlash} and \emph{Carnot} in terms of speed. \emph{NN-PTFlash} is \emph{PTFlash} accelerated by neural networks, as presented in Section \ref{sec:deep_learning}.}
	\label{fig:compare_flash}
\end{figure}

We compare the execution time of different methods for flash calculations with respect to the workload quantified by the number of samples $n$, as shown in Figures \ref{fig:compare_flash}. Due to GPU memory limitations, the maximum number of samples allowed is 10, 5, and 1 million for the three case studies, respectively. We can see that all three figures exhibit the same behavior. When the workload is relatively low, e.g., $n < 1000$, \textit{Carnot} wins by large margins, and CPU is also preferable based on the fact that \textit{PTFlash} runs much faster on CPU than on GPU. On the one hand, PyTorch has some fixed overhead in the setup of the working environment, e.g., the creation of tensors. On the other hand, when GPU is used, there are some additional costs of CPU-GPU communication and synchronization. When $n$ is small, these overheads dominate. As proof, we can see that the time of \emph{PTFlash} on GPU hardly changes as $n$ varies from 100 to $10^4$. In contrast, the time of \emph{Carnot} is almost proportional to $n$.

As the workload increases, the strength of \textit{PTFlash} on GPU emerges and becomes increasingly prominent. For the three case studies, \emph{PTFlash} on GPU is 163.4 (2 components), 106.3 (4 components) and 50.5 (9 components) times faster than \emph{Carnot} at the maximum number of samples. This suggests that \textit{PTFlash} on GPU is more suitable for large-scale computation. Interestingly, we can observe that \textit{PTFlash} on CPU also outperforms \textit{Carnot} when the workload is relatively heavy, e.g., $n > 10^3$. In fact, thanks to Advanced Vector Extensions, vectorization enables fuller utilization of CPU's computational power.

\begin{figure}[h]
	\centering
	\subfigure[Successive substitution]{
		\includegraphics[width=0.47\textwidth]{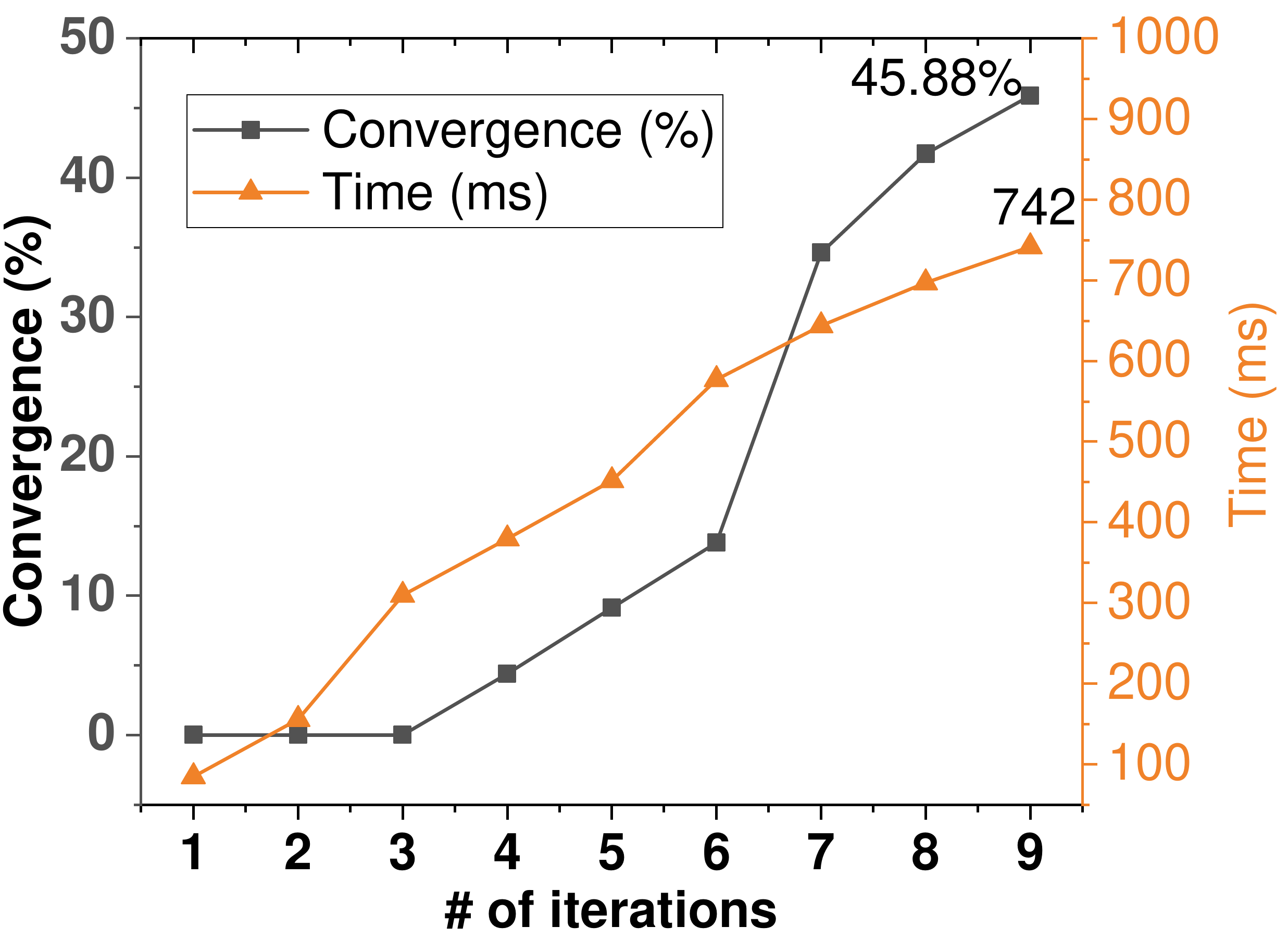}
		\label{fig:ss_splitter}
	}
	\subfigure[Trust-region method]{
		\includegraphics[width=0.47\textwidth]{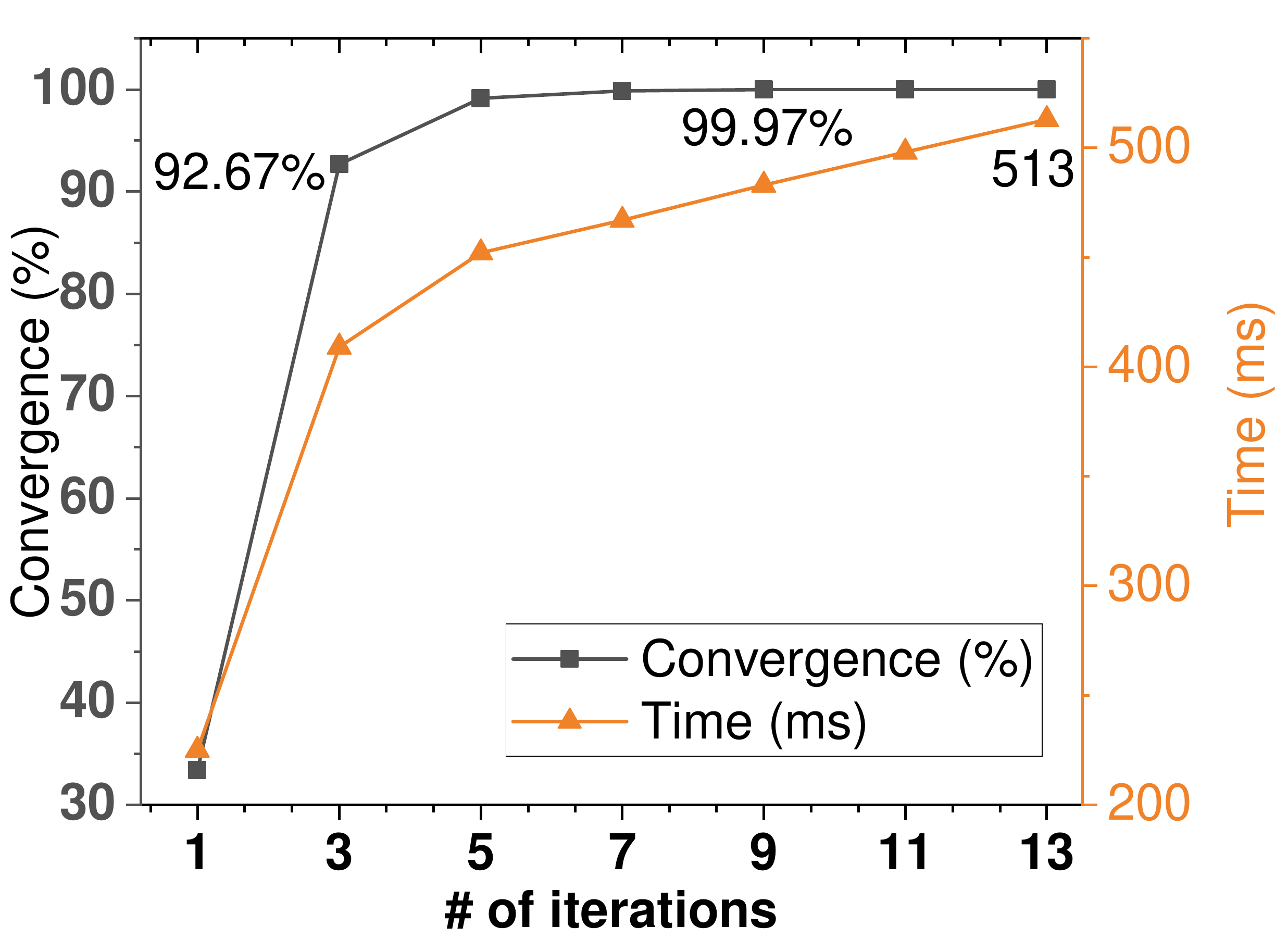}
		\label{fig:tr_splitter}
	}
	\subfigure[Closeness to critical points]{
		\includegraphics[width=0.44\textwidth]{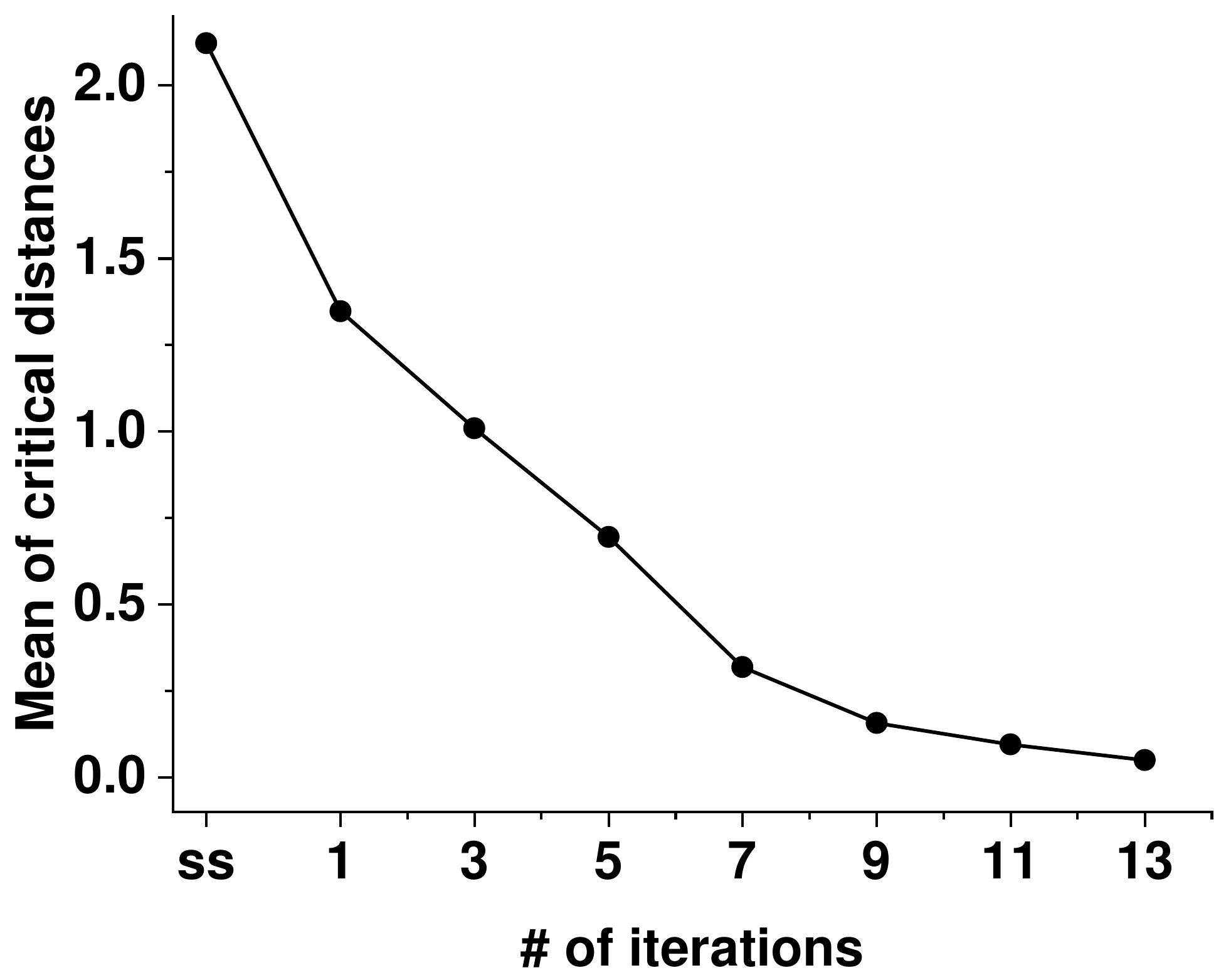}
		\label{fig:cd}
	}
	\caption{Figures (a) and (b) show the convergence percentage and the elapsed time up to each iteration of phase split calculations of \emph{PTFlash} on GPU. In Figure (c), on the x-axis, ss corresponds to the end of successive substitution and other integers are the number of iterations of the trust-region method.}
	\label{fig:splitter}
\end{figure}

Next, we focus on the mixture of 9 components and analyze the performance of \emph{PTFlash} for this case study. Table \ref{tab:profiler} is a performance profiler of \textit{PTFlash} on GPU at $n = 10^6$, which records the running time of each subroutine of flash calculations. As a complement, Figures \ref{fig:splitter} dissect phase split calculations by tracking the total elapsed time and the convergence percentage up to each iteration, as well as the mean of critical distances $d_c$ of converged samples at each iteration, where $d_c$ is defined as:
\begin{equation}
	d_c = \sqrt{\sum_{i=1}^{N_c} \ln K_i^2}  \label{eq:dc}
\end{equation}
The closer to critical points, the smaller $d_c$. In other words, $d_c$ indicates the closeness to critical points. 

The observations of Figures \ref{fig:splitter} are summarized as follows: (1) In Figure \ref{fig:ss_splitter}, the slope of time with respect to the number of iterations is decreasing because the workload is reduced due to incremental convergence. (2) In Figure \ref{fig:tr_splitter}, for the samples that do not converge after successive substitution, the majority of them (92.67\%) converge after 3 iterations of the trust-region method. (3) In Figure \ref{fig:cd}, $d_c$ decreases during iterations, which means that samples close to critical points converge last and also confirms that convergence is slow around critical points.

\begin{table}[h]
	\centering
	\begin{threeparttable}[b]
	\footnotesize
	\caption{Performance profiler of \emph{PTFlash} on GPU (Figure \ref{fig:isothermal_flash}) for the mixture of 9 components at $n = 10^6$ in Figure \ref{fig:compare_9_components}}
	\begin{tabular}{@{}c|c|cccc|cc@{}}
		\toprule
		& \multicolumn{1}{c|}{ss of}                                   & \multicolumn{4}{c|}{Stability analysis}                              & \multicolumn{2}{c}{Phase split}   \\ \cmidrule{3-6}
		& \multicolumn{1}{c|}{phase split}    & \multicolumn{2}{c|}{vapor-like estimate}       & \multicolumn{2}{c|}{liquid-like estimate}    & \multicolumn{2}{c}{calculations}  \\ \cmidrule{7-8}
		& \multicolumn{1}{c|}{calculations}   & ss                 & \multicolumn{1}{c|}{tr}   & ss                   & \multicolumn{1}{c|}{tr}   & ss                                        & tr      \\ \midrule
		\verb|#| of samples                   & $10^6$             & 625645                    & 130715               & 625645                    & 90179               & 413442              & 223741  \\ [1.0ex]
		Convergence                           & 37.44\% \tnote{1}  & 79.11\%                   & 100\%                & 85.59\%                   & 100\%               & 45.88\%             & 100\%   \\ [1.0ex]
		Max number                            & \multirow{2}{*}{3} & \multirow{2}{*}{9}        & \multirow{2}{*}{18}  & \multirow{2}{*}{9}        & \multirow{2}{*}{16} & \multirow{2}{*}{9}  & \multirow{2}{*}{13} \\
		of iterations                         &                    &                           &                      &                           &                     &                     &         \\   [1.0ex]
		\multirow{2}{*}{Total time}           & \multirow{2}{*}{0.4565s}  & 0.4136s            & 0.3417s              & 0.4044s                   & 0.2706s             & 0.7412s             & 0.5132s \\
		                                      &                    &   \multicolumn{4}{c|}{1.3237s \tnote{2}}                                               & \multicolumn{2}{c}{1.2544s}    \\       \midrule\midrule
		\multicolumn{4}{c}{ss: successive substitution}   & \multicolumn{4}{c}{tr: trust-region method} \\
		\bottomrule
	\end{tabular}
	\label{tab:profiler}
	\begin{tablenotes}
		\item [1] 37.44\% is the percentage of samples for which any of $\upDelta G$, $tpd_x$ and $tpd_y$ is negative after 3 attempts of successive substitution, as described in Section \ref{sec:strategy}.
		\item [2] The total time of stability analysis is less than the sum of the times of all subroutines because vapor-like and liquid-like estimates are handled concurrently.
	\end{tablenotes}
\end{threeparttable}
\end{table}

The above analysis gives us a general understanding of \emph{PTFlash}, but in fact it is not easy to analyze \emph{PTFlash} comprehensively because each subroutine also contains iterative algorithms, such as solving the SRK equation of state and the Rice-Rachford equation. Nevertheless, given the information already obtained, we know that we need to shorten the time of stability analysis and reduce the number of iterations in order to accelerate \emph{PTFlash}, which is exactly the role of the classifier and initializer.

\subsection{Deep-learning-powered vectorized flash calculations}

We trained neural networks following Section \ref{sec:deep_learning} for the mixture of 9 components. Here, we will explore the effect of neural networks. First of all, we set $p_l = 0.02$ and $p_r = 0.98$ as the thresholds of stability and instability, which are carefully chosen so that no misclassification occurs. In Figure \ref{fig:compare_9_components}, we can see that \emph{NN-PTFlash} outpaces \emph{PTFlash} on both CPU (2.7x speed-up) and GPU (2.2x speed-up). In addition, \emph{NN-PTFlash} on GPU runs almost 110.7 times faster than \emph{Carnot} at $n = 10^6$.

Table \ref{tab:profiler_nn_flash} is the performance profiler of \emph{NN-PTFlash} on GPU. We can see that the classifier is able to precisely determine the stability of the vast majority of samples (99.42\%), which significantly relieves the burden of stability analysis and saves time. In addition, compared to phase split calculations of \emph{PTFlash}, the convergence percentage of successive substitution increases from 45.88\% to 67.40\%, and the overall time is also greatly reduced, which is contributed to better initial $K_i$ provided by the initializer.

\begin{table}[h]
	\centering
	\begin{threeparttable}[b]
		\footnotesize
		\caption{Performance profiler of \emph{NN-PTFlash} on GPU (Figure \ref{fig:dl_vectorized_flash}) for the mixture of 9 components at $n = 10^6$ in Figure \ref{fig:compare_9_components}}
		\begin{tabular}{@{}c|c|cccc|cc@{}}
			\toprule
			& \multicolumn{1}{c|}{\multirow{3}{*}{classifier}}                                   & \multicolumn{4}{c|}{Stability analysis}             & \multicolumn{2}{c}{Phase split}   \\ \cmidrule{3-6}
			& \multicolumn{1}{c|}{}   & \multicolumn{2}{c|}{vapor-like estimate}        & \multicolumn{2}{c|}{liquid-like estimate}    & \multicolumn{2}{c}{calculations}                  \\ \cmidrule{7-8}
			& \multicolumn{1}{c|}{}   & ss                 & \multicolumn{1}{c|}{tr}    & ss                    & \multicolumn{1}{c|}{tr}   & ss                               & tr        \\ \midrule
			\verb|#| of samples              & $10^6$             & 5818                & 1073                 & 5818                & 1704                & 413442     & 134786           \\  [1.0ex]
			Convergence                      & 99.42\% \tnote{1}  & 81.56\%             & 100\%                & 70.71\%             & 100\%               & 67.40\%    & 100\%            \\  [1.0ex]
			Max number                       &                    & \multirow{2}{*}{9}  & \multirow{2}{*}{13}  & \multirow{2}{*}{9}  & \multirow{2}{*}{12} & \multirow{2}{*}{9}  &   \multirow{2}{*}{13}   \\
			of iterations                    &                    &                     &                      &                     &                     &                     &         \\    [1.0ex]
			\multirow{2}{*}{Total time}      & \multirow{2}{*}{0.0005s}  & 0.1365s      & 0.128s               & 0.0514s             & 0.12s               & 0.7043s             & 0.3388s \\
			                                 &                           &   \multicolumn{4}{c|}{0.34s}                                                    &   \multicolumn{2}{c}{1.0431s} \\  \midrule\midrule
			\multicolumn{4}{c}{ss: successive substitution}   & \multicolumn{4}{c}{tr: trust-region method} \\
			\bottomrule
		\end{tabular}
		\label{tab:profiler_nn_flash}
		\begin{tablenotes}
			\item [1] 99.42\% includes 58.38\% predicted as stable (i.e., $p > p_r$) and 41.04\% predicted as unstable (i.e., $p < p_l$).
		\end{tablenotes}
	\end{threeparttable}
\end{table}

We also performed ablation studies to compare the contributions of the classifier and initializer by using them individually. For instance, when handling 1 million samples for the case study containing 9 components, \emph{NN-PTFlash} with only the classifier on GPU takes 1.88s. However, the attempt to use the initializer alone fails because we found its outputs may reach unreasonably large values (e.g., 1.0$e$15) for stable mixtures far away from the boundary between the single-phase and two-phase regions, which leads to numerical overflow. From machine learning terminology, this is the out-of-distribution generalization problem, since the initializer is trained on the two-phase region and may suffer from large predictive errors when used within the single-phase region. Nonetheless, there is no problem when the initializer works in tandem with the classifier because remaining samples located in the single-phase region are fairly close to the boundary after filtering through the classifier, as shown in Figure \ref{fig:classifier_contour}. In any case, based on the fact that \emph{NN-PTFlash} using only the classifier always lags behind that using both, we can conclude that both the classifier and initializer play an important role in speeding up flash calculations.

\subsection{Discussion}
The results show that the systematic and exhaustive vectorization of isothermal two-phase flash calculations does result in attractive speed-up when large-scale computation is involved, e.g., the number of samples to process is on the order of millions. Importantly, this speed-up does not come at the cost of accuracy and stability like \cite{gaganis_integrated_2014, gaganis_machine_2012, kashinath_fast_2018, wang_accelerating_2019} which are subject to the unreliability of machine learning models. In addition, we can see that neural networks, such as the classifier and initializer, really make a big difference.

Due to GPU memory limitations, the number of samples $n$ is limited in Figures \ref{fig:compare_flash}. Nonetheless, we can see that the slopes of time with respect to $n$ differ significantly between different methods. The time of \emph{Carnot} is proportional to $n$, in contrast, the time of \emph{PTFlash} on GPU is increasing slowly. Therefore, it is reasonable to believe that the speed advantage of \emph{PTFlash} on GPU will become increasingly prominent if $n$ continues to grow. 

Using PyTorch has several benefits in addition to its simplicity and flexibility. First, we can seamlessly incorporate neural networks into \emph{PTFlash}. Second, any subroutine of \emph{PTFlash} is fully differentiable through automatic differentiation, and we can also leverage the implicit function theorem for efficient differentiation, as we did in Section \ref{sec:training_initializer}. Third, PyTorch's highly optimized and ready-to-use multi-GPU parallelization largely circumvents the painstaking hand-crafted effort.

\emph{PTFlash} also has several limitations. First, \emph{PTFlash} is based on the SRK equation of state, which is relatively simple and sufficient for mixtures containing hydrocarbons and non-polar components, but does not take into account the effect of hydrogen bonding and falls short of adequacy for cross-associating mixtures having polar components, such as water and alcohol \cite{kontogeorgis_thermodynamic_2009}. In this case, more advanced but also more complicated equations of state should be employed, such as the SAFT equation of state \cite{wertheim_fluids_1984, wertheim_fluids_1984-1, wertheim_fluids_1986, wertheim_fluids_1986-1, chapman_new_1990, huang_equation_1990} or the CPA equation of state \cite{kontogeorgis_equation_1996, kontogeorgis_multicomponent_1999}. Second, \emph{PTFlash} consumes a large amount of GPU memory, badly limiting its use on much larger batches of data. We need to optimize \emph{PTFlash} to reduce the consumption of GPU memory, e.g., by leveraging the sparsity and symmetry of matrices. Third, \emph{PTFlash} does not support multi-phase equilibrium. Last but not least, neural networks are subject to the out-of-distribution generalization problem. If pressure and temperature are out of predefined ranges used to train neural networks, predictive performance will deteriorate dramatically. Furthermore, once the components of the mixture change, we need to create new neural networks and train them from scratch.

%% file: Sections/conclusion.tex
\section{Conclusion} \label{sec:conclusion}

In this work, we presented a fast and parallel framework, \emph{PTFlash}, for isothermal two-phase flash calculations based on PyTorch and powered by GPU, which efficiently vectorizes algorithms and gains attractive speed-up at large-scale calculations. Two neural networks were used to predict the stability of given mixtures and to initialize the distribution coefficients more accurately than the Wilson approximation, which greatly accelerate \emph{PTFlash}. In addition, \emph{PTFlash} has much broader utility compared to the aforementioned methods which are mainly tailored to compositional reservoir simulation.

We compared \emph{PTFlash} with \emph{Carnot}, an in-house thermodynamic library, and we investigated three case studies containing 2, 4 and 9 components with maximum number of samples of 10, 5 and 1 million, respectively. The results showed that \emph{PTFlash} on GPU is 163.4, 106.3 and 50.5 times faster than \emph{Carnot} at the maximum number of samples for these three cases, respectively.

In the future, we will optimize \emph{PTFlash} to reduce the consumption of GPU memory and extend our work to support multi-phase equilibrium and more advanced equations of state. In addition, we will also apply our work to downstream applications, e.g., compositional reservoir simulation.

%% file: Appendices/srk_eos.tex
\section{SRK equation of state and its solution}  \label{sec:srk_eos}
The SRK equation of state describes the relationship between pressure ($P$), temperature ($T$) and volume ($V$) in the following mathematical form:
\begin{equation}
	P = \frac{RT}{V - b} - \frac{a \alpha}{V (V + b)}  \label{eq:srk}
\end{equation}
where $R$ is the gas constant, $a \alpha$ refers to the temperature-dependent energy parameter, and $b$ denotes the co-volume parameter. We employ the van der Waals mixing rules and the classical combining rules to calculate $a \alpha$ and $b$, as follows:
\begin{subequations} 
	\label{eq:mixing_rule}
	\begin{align}
		a \alpha &= \sum_{i=1}^{N_c} \sum_{j=1}^{N_c} c_i c_j (a \alpha)_{ij}                                      \label{eq:a}    \\
		(a \alpha)_{ij} &= (1 - k_{ij}) \sqrt{(a \alpha)_i (a \alpha)_j}                            \label{eq:a_ij} \\
		b &= \sum_{i=1}^{N_c} c_i b_i                                                                   \label{eq:b}    \\
		a_i &= \frac{0.42748 \cdot R^2 \ (T_{c, i})^2}{P_{c, i}}                                    \label{eq:a_i}  \\
		b_i &= \frac{0.08664 \cdot R \ T_{c, i}}{P_{c, i}}                                          \label{eq:b_i}  \\
		\alpha_i &= \left[ 1 + m_i \left( 1 - \sqrt{\frac{T}{T_{c,i}}} \right) \right]^2        \label{eq:alpha_i}  \\
		m_i &= 0.480 + 1.574 \ \omega_i - 0.176 \ \omega_i^2                                        \label{eq:m_i}
	\end{align}
\end{subequations}
where the subscripts $i$ and $j$ refer to the components $i$ and $j$, respectively, $c_i$ denotes the mole fraction of the component $i$ in the phase considered, $k_{ij}$ is the binary interaction parameter between the components $i$ and $j$, $a_i$ and $b_i$ are two substance-specific constants related to the critical temperature $T_{c, i}$ and critical pressure $P_{c, i}$, and $\omega_i$ is the acentric factor. We reformulate Equation \ref{eq:srk} as a cubic equation in terms of the compressibility factor $Z$:
\begin{equation}
	f_{srk}(Z) = Z^3 - Z^2 + \rho_1 Z - \rho_0 = 0   \label{eq:cubic_Z}
\end{equation}
where $\rho_0 = A B$ and $\rho_1 = A - B (1 + B)$, in which $A = {a \alpha P}/(R^2 T^2)$ and $B = {b P}/(R T)$.
To find the roots of $f_{srk}(Z)$, we utilize an iterative approach based on Halley's method as follows:
\begin{equation}
	Z^{(k+1)} = Z^{(k)} - \frac{f_{srk}(Z^{(k)})}{f_{srk}'(Z^{(k)})} \left[ 1 - \frac{f_{srk}(Z^{(k)})}{f_{srk}'(Z^{(k)})} \cdot \frac{f_{srk}''(Z^{(k)})}{2 f_{srk}'(Z^{(k)})} \right]^{-1}  \label{eq:halley}
\end{equation}

The above iteration starts with a liquid-like guess and converges to a real root $Z_0$ (The convergence criterion is $\lvert Z^{(k+1)} / Z^{(k)} - 1 \rvert <$1.0$e$-8), and then we deflate the cubic equation as:
\begin{equation}
	f_{srk}(Z) = (Z - Z_0)(Z^2 + pZ + q) = 0
\end{equation}
where $p = Z_0 - 1$ and $q = p Z_0 + \rho_1$. If $p^2 < 4q$, only one real root $Z_0$ exists, otherwise, there are three real roots and the other two are $-p/2 \pm \sqrt{p^2 - 4q} / 2$. In the latter case, we assign the smallest root to the liquid phase and the biggest one to the vapour phase. Subsequently, the root corresponding to the lowest Gibbs energy will be chosen. When $Z$ is known, the fugacity coefficients ${\varphi}_i$ are calculated as follows:
\begin{align}
	\ln {\varphi}_i (P, T, \vec{c}) = & \frac{b_i}{b} (Z - 1) - \ln (Z - B) \nonumber \\
	                                  &  + \frac{A}{B} \left( \frac{b_i}{b} - \frac{2}{a \alpha} \sum_{j=1}^{N_c} (a \alpha)_{ij} c_j \right) \ln (1 + \frac{B}{Z})
	\label{eq:lnphi}
\end{align}
where $\vec{c}$ is the composition of the phase considered. In addition, the derivatives of the fugacity coefficients with respect to mole numbers, which are necessary for the trust-region methods of stability analysis and phase split calculations, are calculated explicitly rather than through PyTorch's automatic differentiation, which requires retaining intermediate results and consumes prohibitive memory at large-scale computation.

%% file: Appendices/trust_region_methods.tex
\section{Trust-region method}  \label{sec:tr_method}

When successive substitution fails to converge quickly, particularly around critical points for which liquid and vapor phases are almost indistinguishable, we will switch to the trust-region method with restricted steps, which is a second-order optimization technique, to achieve faster convergence.

\subsection{Trust-region method for stability analysis}  \label{sec:tr_sa}

The objective function to be minimized is the modified tangent plane distance:
\begin{equation}
	tm(\vec{W}) = \sum_{i=1}^{N_c} W_i (\ln W_i + \ln \varphi_i(\vec{W}) - \ln z_i - \ln \varphi_i(\vec{z}) - 1)  \nonumber
\end{equation}

The minimization is accomplished by iterating the following equations:
\begin{subequations}
	\label{eq:newton_analyser}
	\begin{gather}
		\vec{\beta}^{(k)} = 2 \sqrt{\vec{W}^{(k)}}  \\
		(\mat{H}^{(k)} + \eta^{(k)} \mat{I}) \cdot \upDelta \vec{\beta} + \vec{g}^{(k)} = \vec{0} \quad  s.t. \quad \lVert \upDelta \vec{\beta} \rVert \le \upDelta_{max}^{(k)}  \\
		\vec{\beta}^{(k+1)} = \vec{\beta}^{(k)} + \upDelta \vec{\beta}  \\
		\vec{W}^{(k+1)} = \left( \frac{\vec{\beta}^{(k+1)}}{2} \right)^2
	\end{gather}
	\label{eq:tr_sa}
\end{subequations}

where $\mat{I}$ is the identity matrix, $\vec{g}$ and $\mat{H}$ are the gradient and Hessian matrix of $tm$ with respect to $\vec{\beta}$, respectively, and are calculated as follows:

\begin{subequations}
	\begin{gather}
		g_i = \sqrt{W_i} (\ln W_i + \ln \varphi_i(\vec{W}) - \ln z_i - \ln \varphi_i(\vec{z}))  \label{eq:analysis_grad}  \\
		H_{ij} = \sqrt{W_i W_j} \frac{\partial \ln \varphi_i}{\partial W_i} + \sigma_{ij} \left( 1 + \frac{g_i}{\beta_i} \right) \quad \mathrm{where} \ \sigma_{ij} = 1 \ \Leftrightarrow \ i=j \label{eq:analysis_hess}
	\end{gather}
\end{subequations}

In addition, $\eta$ is the trust-region size used to guarantee the positive definiteness of $\mat{H} + \eta \mat{I}$ and to tailor the step size to meet $\lVert \upDelta \vec{\beta} \rVert \le \upDelta_{max}$, where $\upDelta_{max}$ is adjusted during iterations depending on the match between the actual reduction $\delta_{tm} = {tm}^{(k+1)} - {tm}^{(k)}$ and the predicted reduction based on the quadratic approximation $\hat{\delta}_{tm} = \Delta \vec{\beta}^T \vec{g} +  \frac{1}{2} \Delta \vec{\beta}^T \mat{H} \Delta \vec{\beta}$, using the following heuristic rules:
\begin{equation}
	\Delta_{max}^{(k+1)} =
	\begin{cases}
		\dfrac{\Delta_{max}^{(k)}}{2}, & \text{if } \left| \delta_{tm} / \hat{\delta}_{tm} \right| \le 0.25 \vspace{0.6ex} \\
		2 \Delta_{max}^{(k)},          & \text{if } \left| \delta_{tm} / \hat{\delta}_{tm} \right| \ge 0.75 \vspace{0.6ex} \\
		\Delta_{max}^{(k)},            & \text{otherwise}
	\end{cases}
\end{equation}
The convergence criterion of Equations \ref{eq:tr_sa} is $\max ( \lvert \vec{g} \rvert ) <$1.0$e$-6. 

\subsection{Trust-region method for phase split calculations}  \label{sec:tr_split}

The objective function to be minimized is the reduced Gibbs energy:
\begin{equation}
	G = \sum_{i=1}^{N_c} n_i^L (\ln x_i + \ln \varphi_i^L) + \sum_{i=1}^{N_c} n_i^V (\ln y_i + \ln \varphi_i^V) \nonumber
\end{equation}
where $n_i^L = x_i (1 - \theta_V)$ and $n_i^V = y_i \theta_V$ are the mole numbers of liquid and vapor phases, respectively. We choose $n_i^V$ as the independent variable and perform the following iteration:
\begin{subequations}
	\label{eq:newton_splitter}
	\begin{gather}
		\left( \mat{\tilde{H}}^{(k)} + \tilde{\eta}^{(k)} \cdot \vec{D} \left( \frac{\mat{z}}{\mat{x} \mat{y}} \right) \right) \cdot \upDelta \vec{n}^V + \vec{\tilde{g}}^{(k)} = \vec{0} \quad  s.t. \quad \lVert \upDelta \vec{n}^V \rVert \le \tilde{\upDelta}_{max}^{(k)}  \\
		\vec{n}^{V, k+1} = \vec{n}^{V, k} + \upDelta \vec{n}^V
	\end{gather}
\end{subequations}
where $\mat{\tilde{H}}^{(k)}$ and $\vec{\tilde{g}}^{(k)}$ are the gradient and hessian matrix of $G$ with respect to $n_i^V$, respectively, and are calculated as follows:
\begin{subequations}
	\begin{gather}
		\tilde{g}_i = \ln y_i + \ln \varphi_i^V - \ln x_i - \ln \varphi_i^L  \label{eq:split_grad}  \\
		\tilde{H}_{ij} = \frac{1}{\theta_V (1 - \theta_V)} \left( \frac{z_i}{x_i y_i} \sigma_{ij} - 1 + \theta_V \frac{\partial \ln \varphi_i^L}{\partial n_j^L} + (1 - \theta_V) \frac{\partial \ln \varphi_i^V}{\partial n_j^V} \right) \label{eq:split_hess}
	\end{gather}
\end{subequations}

In addition, $\vec{D}(\cdot)$ is a diagonal matrix with diagonal entries in parentheses. The above iteration stops if $\max ( \lvert \vec{\tilde{g}} \rvert ) <$1.0$e$-8. Here, the trust-region method is implemented in the same way as in stability analysis. 

%% file: Appendices/solve_rr_equation.tex
\section{Solution of the Rachford-Rice equation}  \label{sec:solve_RR}

The Rachford-Rice equation is as follows:
\begin{equation}
	f_{RR}(\theta_V, \vec{K}) = \sum_{i=1}^{N_c} \frac{(K_i - 1) z_i}{1 + (K_i - 1) \theta_V} = 0 \nonumber
\end{equation}
Given $\vec{K}$, the solution of the above equation amounts to finding an appropriate zero yielding all non-negative phase compositions. Concretely, we adopt the method proposed by \cite{leibovici_new_1992}, which transforms $f_{RR}$ into a helper function $h_{RR}$ which is more linear in the vicinity of the zero:
\begin{equation}
	h_{RR}(\theta_V, \vec{K}) = (\theta_V - \alpha_l) \cdot (\alpha_r - \theta_V) \cdot f_{RR}(\theta_V) = 0 \label{eq:RR_helper}
\end{equation}
where $\alpha_l = 1 / (1-\max(K_i))$ and $\alpha_r = 1 / (1-\min(K_i))$. The above equation is solved by alternating between the Newton method and the bisection method used when the Newton step renders $\theta_V$ out of the bounds which contain the zero and become narrower during iterations. When the Newton step size is smaller than 1.0$e$-8, the iteration stops.

%% file: Appendices/typical_compositions.tex
\section{Some typical reservoir fluid compositions}  \label{sec:fixed_compositions}

\begin{table}[htbp]
	\centering
	\small
	\caption{Some typical reservoir fluid compositions}
	\begin{tabular}{ccccc}
		\toprule
		& Wet gas & Gas condensate & Volatile oil & Black oil \\
		\midrule
		$CH_4$          & 92.46\% & 73.19\% & 57.6\%  & 33.6\% \\
		$C_2H_6$        & 3.18\%  & 7.8\%   & 7.35\%  & 4.01\% \\
		$C_3H_8$        & 1.01\%  & 3.55\%  & 4.21\%  & 1.01\% \\
		$n$-$C_4H_{10}$ & 0.52\%  & 2.16\%  & 2.81\%  & 1.15\% \\
		$n$-$C_5H_{12}$ & 0.21\%  & 1.32\%  & 1.48\%  & 0.65\% \\
		$C_6H_{14}$     & 0.14\%  & 1.09\%  & 1.92\%  & 1.8\%  \\
		$C_7H_{16}^+$   & 0.82\%  & 8.21\%  & 22.57\% & 57.4\% \\
		$CO_2$          & 1.41\%  & 2.37\%  & 1.82\%  & 0.07\% \\
		$N_2$           & 0.25\%  & 0.31\%  & 0.24\%  & 0.31\% \\
		\bottomrule
	\end{tabular}
	\label{tab:fixed_comp}
\end{table}

%% file: Appendices/vectorized_algorithms.tex
\section{Vectorized algorithms}  \label{sec:alg}

\subsection{Synchronizer}   \label{sec:synchronizer}
\begin{algorithm*}[h]
	\caption{PyTorch pseudo-code of \textit{synchronizer} to save converged results after iteration and remove the corresponding samples}
	\label{alg:synchronizer}
	\DontPrintSemicolon
	\SetNoFillComment
	\SetKwInput{kwInput}{Input}
	\SetKwBlock{kwInit}{Initialization}{end}
	\SetKwBlock{kwSave}{Saving}{end}
	\SetKwBlock{kwRemove}{Removing}{end}
	\SetKwInput{kwOutput}{Output}
	\kwInput{\parbox[t]{0.8\linewidth}{
			Vectorized iterated function $f(\mat{X}, \mathbb{O})$, initial estimate $\mat{X}^{(0)}$, other $f$-related inputs $\mathbb{O}$, convergence criterion $C$, maximum number of iterations $K$}
	}
	\kwInit{
		Set the number of iterations $k \leftarrow 1$                   \;
		Generate a vector $\vec{i}$ containing indices from 0 to $n-1$  \;
		\tcc{$n$ is the number of samples and indexing starts from 0.}
		Create a placeholder matrix $\widetilde{\mat{X}}$ of the same shape as $\mat{X}^{(0)}$
	}
	\While{$k \le K$}{
		$X^{(k+1)} \leftarrow f(X^{(k)}, \ \mathbb{O})$  \;
		mask $ \leftarrow C(\cdots)$ \;
		\tcc{$C$ returns a Boolean vector and True means convergence.}
		\kwSave{
			indices $ \leftarrow \vec{i}[\mathrm{mask}]$  \;
			$\widetilde{\mat{X}}[\mathrm{indices}] \leftarrow \mat{X}^{(k+1)}[\mathrm{mask}]$  \;
		}
		\kwRemove{
			$\vec{i} \leftarrow \vec{i}[\sim \mathrm{mask}]$  \;
			$\mathbb{O} \leftarrow \mathbb{O}[\sim \mathrm{mask}]$ \;
			\tcc{Apply this operation to every element in $\mathbb{O}$}
			$\mat{X}^{(k+1)} \leftarrow \mat{X}^{(k+1)}[\sim \mathrm{mask}]$  \;
		}
		$k \leftarrow k + 1$  \;
	}
	\If{$\mathrm{len}(\vec{i}) \ne 0$}{
		$\widetilde{\mat{X}}[\vec{i}] \leftarrow \mat{X}$  \;
		\tcc{Also save unconverged results for further utilization.}
	}
	\kwOutput{Converged results $\widetilde{\mat{X}}$ and unconverged indices $\vec{i}$}
\end{algorithm*}
\clearpage
\subsection{Vectorized stability analysis}  \label{sec:vectorized_sa}
\begin{algorithm*}[h]
	\caption{PyTorch pseudo-code of vectorized stability analysis}
	\label{alg:analyser}
	\DontPrintSemicolon
	\SetNoFillComment
	\SetKwInput{kwInput}{Input}
	\SetKwBlock{kwInit}{Initialization}{end}
	\SetKwBlock{kwSS}{Successive substitution}{end}
	\SetKwBlock{kwNewton}{Trust-region method}{end}
	\SetKwBlock{Loop}{While}{end}
	\SetKwInput{kwOutput}{Output}
	\kwInput{\parbox[t]{0.8\linewidth}{
			Pressure $\vec{P}$, temperature $\vec{T}$, feed composition $\textbf{z}$, component properties ($\vec{P_c}$, $\vec{T_c}$, $\vec{\omega}$, BIPs), initial estimate $\mat{W}^{(0)}$, convergence criteria $C_{ss}$ and $C_{tr}$, maximum numbers of iterations $K_{ss} = 9$ and $K_{tr} = 20$}
	}
	\kwInit{
		Instantiate \textit{pteos} = \textit{PTEOS}($\vec{P_c}$, $\vec{T_c}$, $\vec{\omega}$, BIPs) \;
		\tcc{\textit{PTEOS} is a PyTorch-based class to efficiently calculate the fugacity coefficients and their partial derivatives.}
	}
	\kwSS{
		Iterated function $f_{ss}$ specified by Equation \ref{eq:ss_analyser}  \;
		Other inputs $\mathbb{O}_{ss} \leftarrow \{\vec{P}$, $\vec{T}$, $\textbf{z}$\}  \;
		$\mat{W}, \vec{i}_{ss} \leftarrow \mathit{synchronizer}(f_{ss}, \mat{W}^{(0)}, \mathbb{O}_{ss}, \mathbb{C}_{ss}, K_{ss})$
	}
	\kwNewton{
		Iterated function $f_{tr}$ specified by Equations \ref{eq:newton_analyser}  \;
		$\mat{W}^{(0)}_{tr} \leftarrow \mat{W}[\vec{i}_{ss}]$  \;
		Other inputs $\mathbb{O}_{tr} \leftarrow \{\vec{P}[\vec{i}_{ss}]$, $\vec{T}[\vec{i}_{ss}]$, $\textbf{z}[\vec{i}_{ss}]$\}  \;
		$\mat{W}_{tr}, \vec{i}_{tr} \leftarrow \mathit{synchronizer}(f_{tr}, \mat{W}^{(0)}_{tr}, \mathbb{O}_{tr}, \mathbb{C}_{tr}, K_{tr})$ \;
	}
	$\mat{W}[\vec{i}_{ss}] \leftarrow \mat{W}_{tr}$ and $\vec{i} \leftarrow \vec{i}_{ss}[\vec{i}_{tr}]$  \;
	\kwOutput{Converged results $\mat{W}$ and unconverged indices $\vec{i}$}
\end{algorithm*}

%% file: arXiv_PTFlash.bbl
\begin{thebibliography}{10}
\expandafter\ifx\csname url\endcsname\relax
  \def\url#1{\texttt{#1}}\fi
\expandafter\ifx\csname urlprefix\endcsname\relax\def\urlprefix{URL }\fi
\expandafter\ifx\csname href\endcsname\relax
  \def\href#1#2{#2} \def\path#1{#1}\fi

\bibitem{michelsen_isothermal_1982}
M.~L. Michelsen, The isothermal flash problem. part {II}. phase-split
  calculation 9~(1)  21--40, publisher: Elsevier.

\bibitem{michelsen_isothermal_1982-1}
M.~L. Michelsen, The isothermal flash problem. part i. stability 9~(1)  1--19,
  publisher: Elsevier.

\bibitem{wang_non-iterative_1994}
P.~Wang, E.~H. Stenby, Non-iterative flash calculation algorithm in
  compositional reservoir simulation 95  93--108, publisher: Elsevier.

\bibitem{belkadi_comparison_2011}
A.~Belkadi, W.~Yan, M.~L. Michelsen, E.~H. Stenby, Comparison of two methods
  for speeding up flash calculations in compositional simulations, in: {SPE}
  Reservoir Simulation Symposium, {OnePetro}.

\bibitem{dogru_next-generation_2009}
A.~H. Dogru, L.~S.~K. Fung, U.~Middya, T.~Al-Shaalan, J.~A. Pita, A
  next-generation parallel reservoir simulator for giant reservoirs, in: {SPE}
  Reservoir Simulation Symposium, {OnePetro}.

\bibitem{michelsen_simplified_1986}
M.~L. Michelsen, Simplified flash calculations for cubic equations of state
  25~(1)  184--188, publisher: {ACS} Publications.

\bibitem{hendriks_reduction_1988}
E.~M. Hendriks, Reduction theorem for phase equilibrium problems 27~(9)
  1728--1732, publisher: {ACS} Publications.

\bibitem{hendriks_application_1992}
E.~M. Hendriks, A.~Van~Bergen, Application of a reduction method to phase
  equilibria calculations 74  17--34, publisher: Elsevier.

\bibitem{rasmussen_increasing_2006}
C.~P. Rasmussen, K.~Krejbjerg, M.~L. Michelsen, K.~E. Bjurstrøm, Increasing
  the computational speed of flash calculations with applications for
  compositional, transient simulations 9~(1)  32--38, publisher: {OnePetro}.

\bibitem{voskov_compositional_2007}
D.~Voskov, H.~A. Tchelepi, Compositional space parameterization for flow
  simulation, in: {SPE} Reservoir Simulation Symposium, {OnePetro}.

\bibitem{gaganis_integrated_2014}
V.~Gaganis, N.~Varotsis, An integrated approach for rapid phase behavior
  calculations in compositional modeling 118  74--87, publisher: Elsevier.

\bibitem{gaganis_machine_2012}
V.~Gaganis, N.~Varotsis, Machine learning methods to speed up compositional
  reservoir simulation, in: {SPE} Europec/{EAGE} annual conference, {OnePetro}.

\bibitem{kashinath_fast_2018}
A.~Kashinath, M.~Szulczewski, A.~Dogru, A fast algorithm for calculating
  isothermal phase behavior using machine learning 465  73--82.
\newblock \href {https://doi.org/10.1016/j.fluid.2018.02.004}
  {\path{doi:10.1016/j.fluid.2018.02.004}}.

\bibitem{wang_accelerating_2019}
S.~Wang, N.~Sobecki, D.~Ding, L.~Zhu, Y.-S. Wu, Accelerating and stabilizing
  the vapor-liquid equilibrium ({VLE}) calculation in compositional simulation
  of unconventional reservoirs using deep learning based flash calculation 253
  209--219.
\newblock \href {https://doi.org/10.1016/j.fuel.2019.05.023}
  {\path{doi:10.1016/j.fuel.2019.05.023}}.

\bibitem{cortes_support-vector_1995}
C.~Cortes, V.~Vapnik, Support-vector networks 20~(3)  273--297, publisher:
  Springer.

\bibitem{tipping_sparse_2001}
M.~E. Tipping, Sparse bayesian learning and the relevance vector machine 1
  211--244.

\bibitem{goodfellow_deep_2016}
I.~Goodfellow, Y.~Bengio, A.~Courville, Deep learning, {MIT} press.

\bibitem{chen_gpu-based_2014}
Z.~Chen, H.~Liu, S.~Yu, B.~Hsieh, L.~Shao, {GPU}-based parallel reservoir
  simulators, in: Domain Decomposition Methods in Science and Engineering
  {XXI}, Springer, pp. 199--206.

\bibitem{soave_equilibrium_1972}
G.~Soave, Equilibrium constants from a modified redlich-kwong equation of state
  27~(6)  1197--1203, publisher: Elsevier.

\bibitem{paszke_pytorch_2019}
A.~Paszke, S.~Gross, F.~Massa, A.~Lerer, J.~Bradbury, G.~Chanan, T.~Killeen,
  Z.~Lin, N.~Gimelshein, L.~Antiga, {others}, Pytorch: An imperative style,
  high-performance deep learning library 32  8026--8037.

\bibitem{lomont_introduction_2011}
C.~Lomont, Introduction to intel advanced vector extensions 23.

\bibitem{sanders_cuda_2010}
J.~Sanders, E.~Kandrot, {CUDA} by example: an introduction to general-purpose
  {GPU} programming, Addison-Wesley Professional.

\bibitem{zhi_fallibility_2002}
Y.~Zhi, H.~Lee, Fallibility of analytic roots of cubic equations of state in
  low temperature region 201~(2)  287--294, publisher: Elsevier.

\bibitem{orbach_convergence_1971}
O.~Orbach, C.~Crowe, Convergence promotion in the simulation of chemical
  processes with recycle-the dominant eigenvalue method 49~(4)  509--513,
  publisher: Wiley Online Library.

\bibitem{hebden_algorithm_1973}
M.~Hebden, An algorithm for minimization using exact second
  derivativesPublisher: Citeseer.

\bibitem{rachford_procedure_1952}
H.~H. Rachford, J.~Rice, Procedure for use of electronic digital computers in
  calculating flash vaporization hydrocarbon equilibrium 4~(10)  19--3,
  publisher: {OnePetro}.

\bibitem{leibovici_new_1992}
C.~Leibovici, J.~Neoschil, A new look at the rachford-rice equation 74
  303--308.
\newblock \href {https://doi.org/10.1016/0378-3812(92)85069-K}
  {\path{doi:10.1016/0378-3812(92)85069-K}}.

\bibitem{michelsen_thermodynamic_2004}
M.~L. Michelsen, J.~Mollerup, Thermodynamic modelling: fundamentals and
  computational aspects, Tie-Line Publications.

\bibitem{mckay_comparison_2000}
M.~D. {McKay}, R.~J. Beckman, W.~J. Conover, A comparison of three methods for
  selecting values of input variables in the analysis of output from a computer
  code 42~(1)  55--61, publisher: Taylor \& Francis.

\bibitem{bradbury_jax_2018}
J.~Bradbury, R.~Frostig, P.~Hawkins, M.~J. Johnson, C.~Leary, D.~Maclaurin,
  G.~Necula, A.~Paszke, J.~{VanderPlas}, S.~Wanderman-Milne, Q.~Zhang,
  \href{http://github.com/google/jax}{{JAX}: composable transformations of
  python+{NumPy} programs}.
\newline\urlprefix\url{http://github.com/google/jax}

\bibitem{van_rossum_python_2011}
G.~Van~Rossum, F.~L. Drake, The python language reference manual, Network
  Theory Ltd.

\bibitem{hendrycks_gaussian_2016}
D.~Hendrycks, K.~Gimpel, Gaussian error linear units (gelus).

\bibitem{elfwing_sigmoid-weighted_2018}
S.~Elfwing, E.~Uchibe, K.~Doya, Sigmoid-weighted linear units for neural
  network function approximation in reinforcement learning 107  3--11,
  publisher: Elsevier.

\bibitem{ramachandran_swish_2017}
P.~Ramachandran, B.~Zoph, Q.~V. Le, Swish: a self-gated activation function 7
  1, publisher: Technical report.

\bibitem{bergstra_algorithms_2011}
J.~Bergstra, R.~Bardenet, Y.~Bengio, B.~Kégl, Algorithms for hyper-parameter
  optimization 24.

\bibitem{li_system_2020}
L.~Li, K.~Jamieson, A.~Rostamizadeh, E.~Gonina, J.~Ben-Tzur, M.~Hardt,
  B.~Recht, A.~Talwalkar, A system for massively parallel hyperparameter tuning
  2  230--246.

\bibitem{kingma_adam_2014}
D.~P. Kingma, J.~Ba, Adam: A method for stochastic optimization.

\bibitem{smith_no_2015}
L.~N. Smith, No more pesky learning rate guessing games 5.

\bibitem{smith_cyclical_2017}
L.~N. Smith, Cyclical learning rates for training neural networks, in: 2017
  {IEEE} winter conference on applications of computer vision ({WACV}), {IEEE},
  pp. 464--472.

\bibitem{smith_super-convergence_2019}
L.~N. Smith, N.~Topin, Super-convergence: Very fast training of neural networks
  using large learning rates, in: Artificial Intelligence and Machine Learning
  for Multi-Domain Operations Applications, Vol. 11006, International Society
  for Optics and Photonics, p. 1100612.

\bibitem{prechelt_early_1998}
L.~Prechelt, Early stopping-but when?, in: Neural Networks: Tricks of the
  trade, Springer, pp. 55--69.

\bibitem{he_deep_2016}
K.~He, X.~Zhang, S.~Ren, J.~Sun, Deep residual learning for image recognition,
  in: Proceedings of the {IEEE} conference on computer vision and pattern
  recognition, pp. 770--778.

\bibitem{cheng_wide_2016}
H.-T. Cheng, L.~Koc, J.~Harmsen, T.~Shaked, T.~Chandra, H.~Aradhye,
  G.~Anderson, G.~Corrado, W.~Chai, M.~Ispir, {others}, Wide \& deep learning
  for recommender systems, in: Proceedings of the 1st workshop on deep learning
  for recommender systems, pp. 7--10.

\bibitem{krantz_implicit_2002}
S.~G. Krantz, H.~R. Parks, The implicit function theorem: history, theory, and
  applications, Springer Science \& Business Media.

\bibitem{kontogeorgis_thermodynamic_2009}
G.~M. Kontogeorgis, G.~K. Folas, Thermodynamic models for industrial
  applications: from classical and advanced mixing rules to association
  theories, John Wiley \& Sons.

\bibitem{wertheim_fluids_1984}
M.~S. Wertheim, Fluids with highly directional attractive forces. {II}.
  thermodynamic perturbation theory and integral equations 35~(1)  35--47,
  publisher: Springer.

\bibitem{wertheim_fluids_1984-1}
M.~Wertheim, Fluids with highly directional attractive forces. i. statistical
  thermodynamics 35~(1)  19--34, publisher: Springer.

\bibitem{wertheim_fluids_1986}
M.~Wertheim, Fluids with highly directional attractive forces. {IV}.
  equilibrium polymerization 42~(3)  477--492, publisher: Springer.

\bibitem{wertheim_fluids_1986-1}
M.~Wertheim, Fluids with highly directional attractive forces. {III}. multiple
  attraction sites 42~(3)  459--476, publisher: Springer.

\bibitem{chapman_new_1990}
W.~G. Chapman, K.~E. Gubbins, G.~Jackson, M.~Radosz, New reference equation of
  state for associating liquids 29~(8)  1709--1721, publisher: {ACS}
  Publications.

\bibitem{huang_equation_1990}
S.~H. Huang, M.~Radosz, Equation of state for small, large, polydisperse, and
  associating molecules 29~(11)  2284--2294, publisher: {ACS} Publications.

\bibitem{kontogeorgis_equation_1996}
G.~M. Kontogeorgis, E.~C. Voutsas, I.~V. Yakoumis, D.~P. Tassios, An equation
  of state for associating fluids 35~(11)  4310--4318, publisher: {ACS}
  Publications.

\bibitem{kontogeorgis_multicomponent_1999}
G.~M. Kontogeorgis, I.~V. Yakoumis, H.~Meijer, E.~Hendriks, T.~Moorwood,
  Multicomponent phase equilibrium calculations for water–methanol–alkane
  mixtures 158  201--209, publisher: Elsevier.

\end{thebibliography}
